\documentclass[aps,prstab,twocolumn,groupedaddress,showpacs,showkeys]{revtex4-1}
\usepackage{tabularx}
\usepackage{multirow}
\usepackage{amsmath,amssymb}
\usepackage{lmodern}
\usepackage{graphicx}
\usepackage{epstopdf}
\usepackage[colorlinks=true,linkcolor=red,bookmarks=true,citecolor=blue,urlcolor=blue]{hyperref}
\usepackage[T1]{fontenc}
\usepackage{color}

\begin{document}

%\preprint{}

\title{Dielectric laser acceleration of electrons in the vicinity of single and double grating structures - theory and simulations}


\author{John Breuer}
\affiliation{Max Planck Institute of Quantum Optics, Hans-Kopfermann-Str.\ 1, 85748 Garching, Germany, EU}
\email[]{Present Address: Integrated Circuit Testing GmbH,
Ammerthalstra§e 20, 85551 Heimstetten, Germany.}
\author{Joshua McNeur}
\affiliation{Department of Physics, Friedrich-Alexander-Universit{\"a}t Erlangen N{\"u}rnberg, Staudtstr.\ 1, 91058 Erlangen, Germany, EU}
\author{Peter Hommelhoff}
\email[]{peter.hommelhoff@fau.de}
%\thanks{}
\affiliation{Department of Physics, Friedrich-Alexander-Universit{\"a}t Erlangen N{\"u}rnberg, Staudtstr.\ 1, 91058 Erlangen, Germany, EU}
\affiliation{Max Planck Institute of Quantum Optics, Hans-Kopfermann-Str.\ 1, 85748 Garching, Germany, EU}


\date{\today}

\begin{abstract}
Dielectric laser acceleration of electrons close to a fused-silica grating has recently been observed [Peralta \textit{et al.}, Nature 503, 91 (2013); Breuer, Hommelhoff, PRL 111, 134803 (2013)]. Here we present the theoretical description of the near-fields close to such a grating that can be utilized to accelerate non-relativistic electrons. We also show simulation results of electrons interacting with such fields in a single and double grating structure geometry and discuss dephasing effects that have to be taken into account when designing a photonic-structure-based accelerator for non-relativistic electrons. We further model the space charge effect using the paraxial ray equation and discuss the resulting expected peak currents for various parameter sets. % We find that space charge forces will limit the peak current in such devices to $\sim$20\,mA for 30\,keV electrons and $\sim$2\,A for 1\,MeV electrons.
\end{abstract}


\maketitle

%\section{}
%\subsection{}
%\subsubsection{}
\section{Introduction}
The development of lasers emitting pulses with high peak electric fields kindled the vision of a next generation of linear accelerators (linacs) already half a century ago \cite{Shimoda1962}. Modern linac facilities operate with either room temperature or superconducting radio frequency (RF) cavities providing acceleration gradients in the range of 20-50\,MeV/m. Although RF structures with acceleration gradients of up to 100\,MeV/m have been tested \cite{Spataro2011}, the current technology has difficulty reaching higher gradients due to limitations given by breakdown phenomena \cite{Solyak2009}. \textit{Dielectric} materials at \textit{optical} frequencies withstand up to two orders of magnitude larger surface fields than metals \cite{lenzner1998} suggesting the idea of dielectric laser accelerators (DLAs) \cite{Rosenzweig1995,Huang1996}.

Direct acceleration of a charged particle with the electromagnetic carrier field of a laser pulse requires an electromagnetic wave with a phase speed equal to and an electric field component parallel to the particle's velocity in order to continuously impart energy to the particle. Plane waves in vacuum cannot be used for synchronous direct acceleration because they are transversely polarized and propagate at the speed of light \cite{Lawson1979}. Although the longitudinal electric field component inside the focus of a laser beam can be used to directly accelerate particles, the accelerating field propagates faster than the speed of light and is therefore asynchronous with particles interacting with it \cite{Scully1991}; the acceleration distance is limited to approximately the Rayleigh length of the laser focus. Another acceleration scheme that involves lasers is laser-driven plasma-based acceleration \cite{esarey2009}. Here the accelerating fields are provided by a plasma wave, which is excited by short laser pulses.%, and therefore particles are not directly accelerated by the laser field.

Synchronous modes with a longitudinal electric field component can exist close to periodic grating structures, which have been proposed as particle accelerators decades ago \cite{Takeda1968,palmer1980}. Direct acceleration in close vicinity of a grating, also known as the inverse Smith-Purcell effect \cite{Mizuno1975}, has been first observed with a terahertz radiation source at a metal grating with a 250\ $\mu$m period \cite{Mizuno1987,Bae1992}. However, the measured acceleration gradients were too small (keV/m) to compete with conventional RF linacs.

Laser-driven acceleration of relativistic electrons in the optical regime has been first observed at a single matter-vacuum interface which was not a grating \cite{plettner2005b,plettner2005}. The maximum acceleration gradient was 40\,MeV/m for 30\,MeV electrons. But straightforward concatenation of elements, i.e., scalability is technically difficult to achieve with the reported scheme for geometrical reasons.

Plettner \textit{et al.}\ proposed scalable dielectric double grating structures \cite{plettner2006,plettner2008,plettner2008b,plettner2009,Plettner2011}, where electrons propagate in a channel between two gratings facing each other. Accelerating, deflecting and bunching structures can be designed and hence an all-optical dielectric-based table-top accelerator or even free-electron laser seems feasible, despite the small expected bunch charges imposed by the space charge effect and wakefield radiation losses \cite{plettner2008b}. The structures are non-resonating and therefore allow ultrashort-pulsed  ($\sim$10$-$100\,fs) operation. The concept is different from resonating approaches (traveling or standing wave structures) such as photonic bandgap structures \cite{Rosenzweig1995,Yoder2005,Cowan2008,Naranjo2012} that are conceptually similar to conventional RF structures and have filling times on the order of picoseconds.

We have recently observed dielectric laser acceleration of non-relativistic 28\,keV electrons close to a single fused silica grating using the inverse Smith-Purcell effect and measured a maximum acceleration gradient of 25\,MeV/m \cite{Breuer2013}. Together with the concurrent demonstration of dielectric laser acceleration of relativistic 60\,MeV electrons exploiting a fused silica double grating \cite{Peralta2013}, these two experiments prove the concept of dielectric laser acceleration. Their direct intercompatibility bolsters the case for all-optical dielectric accelerators.

Here we investigate the interaction of non-relativistic and moderately relativistic electrons with electromagnetic fields in the vicinity of dielectric single and double grating structures. We present simulations for different grating geometries, estimate the dephasing length and the effects of space charge forces. Finally, we discuss implications of our simulation results for future optical linacs.

\section{Particle acceleration with the electromagnetic field of evanescent waves}

\begin{figure}
	\centering
		\includegraphics[width=\columnwidth]{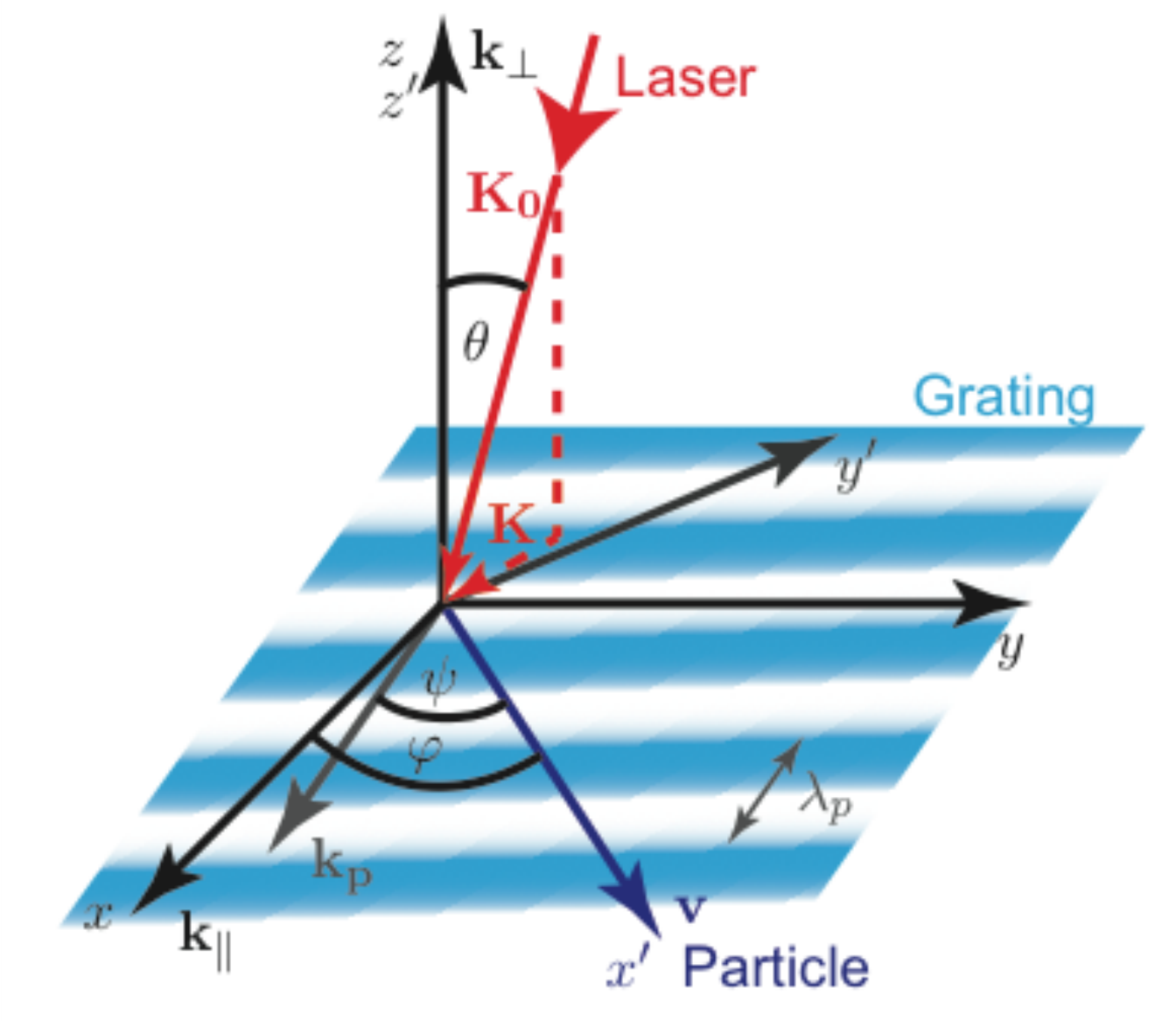}
	\caption{$\{x,y,z\}$ represents the coordinate system for a spatial harmonic, which is excited at an infinitely large grating with grating period $\lambda_\mathrm{p}$ and grating vector $k_\mathrm{p}=2\pi/\lambda_\mathrm{p}$. $\{x',y',z'\}$ is the reference frame for a particle moving parallel to the grating surface, hence $z'=z$. $\psi$ is the angle between $\mathbf{k_\mathrm{p}}$ and the particle's velocity $\mathbf{v}$; $\varphi$ is the angle between the propagation direction $\mathbf{k_\|}$ of the spatial harmonic and $\mathbf{v}$. The wave vector of the incident plane wave is $\mathbf{K_0}$ with the in-plane projection $\mathbf{K}$. The in-plane wave vector of the $n$-th diffracted wave can be written as  $\mathbf{k_\|}^{n}=\mathbf{K}+n\mathbf{k_\mathrm{p}}$ and hence $\varphi$ is determined by $\mathbf{K_0}$ and $\psi$. }
	\label{fig:coord_sys}
\end{figure}

Palmer explored the fields above a single grating excited by a plane wave \cite{palmer1980}. He derived conditions for which particle acceleration with such fields is possible. Following his discussion we analyze the electromagnetic fields close to an infinitely large plane, which is assumed to be a grating with grating period $\lambda_\mathrm{p}$ and $k_\mathrm{p}=2\pi/\lambda_\mathrm{p}$ (see Fig.\ \ref{fig:coord_sys}). The diffraction of the incident wave at the grating excites spatial harmonics with wave vectors $\mathbf{k_\|}^{n}=\mathbf{K}+n\mathbf{k_\mathrm{p}}$, with the in-plane projection of the incident wave vector $\mathbf{K}$ (Fig.\ \ref{fig:coord_sys}) and the order number $n=0,1,2,\dots$\ The electromagnetic field of the $n$-th mode $\mathbf{A}(\mathbf{r},t)=(\mathbf{E}(\mathbf{r},t),\mathbf{B}(\mathbf{r},t))$, with the electric field $\mathbf{E}$ and the magnetic field $\mathbf{B}$, can be written as
\begin{equation}
\mathbf{A}(\mathbf{r},t)=\mathbf{A_n}e^{i(k_\bot^n z+ \mathbf{k_\|^n}\cdot \mathbf{r} -\omega t +\phi)}.
\label{eq:int0}
\end{equation}
Here, $\omega$ is the incident wave's angular frequency and $\phi$ is a phase term. The total field above the grating surface is comprised of a Fourier series of all spatial harmonics.

We assume a single particle traveling parallel to the plane with the trajectory $\mathbf{r}(t)=\mathbf{v}t$, with the velocity $v=\left|\mathbf{v}\right|=\beta c$, at an angle $\varphi$ relative to $\mathbf{k}_\|$ (Fig.\ \ref{fig:coord_sys}). Continuous motional control of the particle requires the component of the accelerating mode's phase velocity, which is parallel to the particle's trajectory, $v_\mathrm{ph}=\omega/(k_\| \cos \varphi)$ to equal $v$. This requirement yields the synchronicity condition
\begin{equation}
k_\|=\frac{\omega}{\beta c \cos \varphi}=\frac{k_0}{\beta \cos \varphi},
\label{eq:int1}
\end{equation}
with the wave vector of the incident plane wave in vacuum $k_0=|\mathbf{K_0}|=\omega/c=2\pi/\lambda$ and wavelength $\lambda$.

In the following we will consider only the synchronous mode for which Eq.\ \ref{eq:int1} is satisfied. We focus on acceleration in vacuum, which implies that the fields have to satisfy the wave equation
\begin{equation}
\left(\nabla^2-\frac{1}{c^2}\partial_t^2\right)\mathbf{A}(\mathbf{r},t)=0.
\label{eq:int2}
\end{equation}
This yields $k_\bot^2+k_\|^2-\omega^2/c^2=0$. Therefore
\begin{equation}
k_\bot=k_0\sqrt{1-\frac{1}{\beta^2\cos^2 \varphi}}=i\frac{k_0}{\widetilde{\beta} \widetilde{\gamma}},
\label{eq:int4}
\end{equation}
with $\widetilde{\beta}=\beta \cos \varphi$ and $\widetilde{\gamma}=\left(1-\widetilde{\beta}^2\right)^{-1/2}$ \cite{palmer1980,Plettner2011}. The accelerating fields fall off exponentially perpendicular to the particle trajectory, since $\beta$\,$<$\,1 and $\gamma$ is real. In other words, just evanescent fields contribute to the acceleration, in agreement with the Lawson-Woodward theorem \cite{Lawson1979,PALMER1988}. Particles have to pass the grating surface within a distance on the order of the (transverse) decay length
\begin{equation}
\delta:=\frac{i}{k_\bot}=\frac{\widetilde{\beta} \widetilde{\gamma} \lambda}{2\pi},
\label{eq:int4b}
\end{equation}
to experience acceleration comparable to the maximum accelerating gradient. Furthermore, synchronous steering (i.e., continuous motional control) of particles with $\beta$\,$\rightarrow$\,0 is virtually impossible with this scheme (since $\delta$\,$\rightarrow$\,0).

We can now calculate the electromagnetic fields of the synchronous mode (with $k_{\|}=k_{x}$) using
\begin{equation}
\mathbf{k}=k_0
\left(\begin{array}{c}
	1/\widetilde{\beta}\\
	0\\
	i/(\widetilde{\beta}\widetilde{\gamma})
\end{array}\right),
\label{eq:int4c}
\end{equation}
\begin{align}
\nabla \times \mathbf{E}&=- \frac{\partial \mathbf{B}}{\partial t},& &\text{and}
&\nabla \times \mathbf{B}&= \frac{1}{c^2} \frac{\partial \mathbf{E}}{\partial t}.
\label{eq:int5}
\end{align}
We obtain
\begin{align}
\mathbf{E}&=
\left(\begin{array}{c}
icB_y/ (\widetilde{\beta}\widetilde{\gamma})\\
E_y\\
-cB_y/\widetilde{\beta}
\end{array}\right),\text{ and }
\mathbf{B}&=
\left(\begin{array}{c}
-iE_y/(\widetilde{\beta} c \widetilde{\gamma})\\
B_y\\
E_y/(\widetilde{\beta} c)
\end{array}\right).
\label{eq:int6}
\end{align}
There are two independent solutions corresponding to the transverse electric (TE) and transverse magnetic (TM) mode. The amplitudes $E_y$ and $B_y$ have to be calculated for each geometry individually.

From the fields we can compute the Lorentz force
\begin{equation}
\begin{split}
\mathbf{F}&=q(\mathbf{E} + \mathbf{v} \times \mathbf{B}) \\
&=q\left(
\begin{array}{c}
icB_y/(\widetilde{\beta} \widetilde{\gamma}) + \tan \varphi E_y\\
0\\
-cB_y(1-\widetilde{\beta}^2)/\widetilde{\beta}+i\tan \varphi E_y /\widetilde{\gamma}
\end{array}\right).
\end{split}
\label{eq:int7}
\end{equation}
Projecting into the particles coordinate system $\{x',y',z'\}$ yields
\begin{equation}
\mathbf{F_{r'}}=q\left(
\begin{array}{c}
icB_y /(\beta \widetilde{\gamma})	+ E_y \sin \varphi\\
-icB_y \tan \varphi /(\beta \widetilde{\gamma})	- E_y \sin \varphi \tan \varphi\\
-cB_y/(\widetilde{\beta} \widetilde{\gamma}^2) + iE_y \tan \varphi /\widetilde{\gamma}
\end{array}\right).
\label{eq:int8}
\end{equation}

\begin{figure}
	\centering
		\includegraphics[width=\columnwidth]{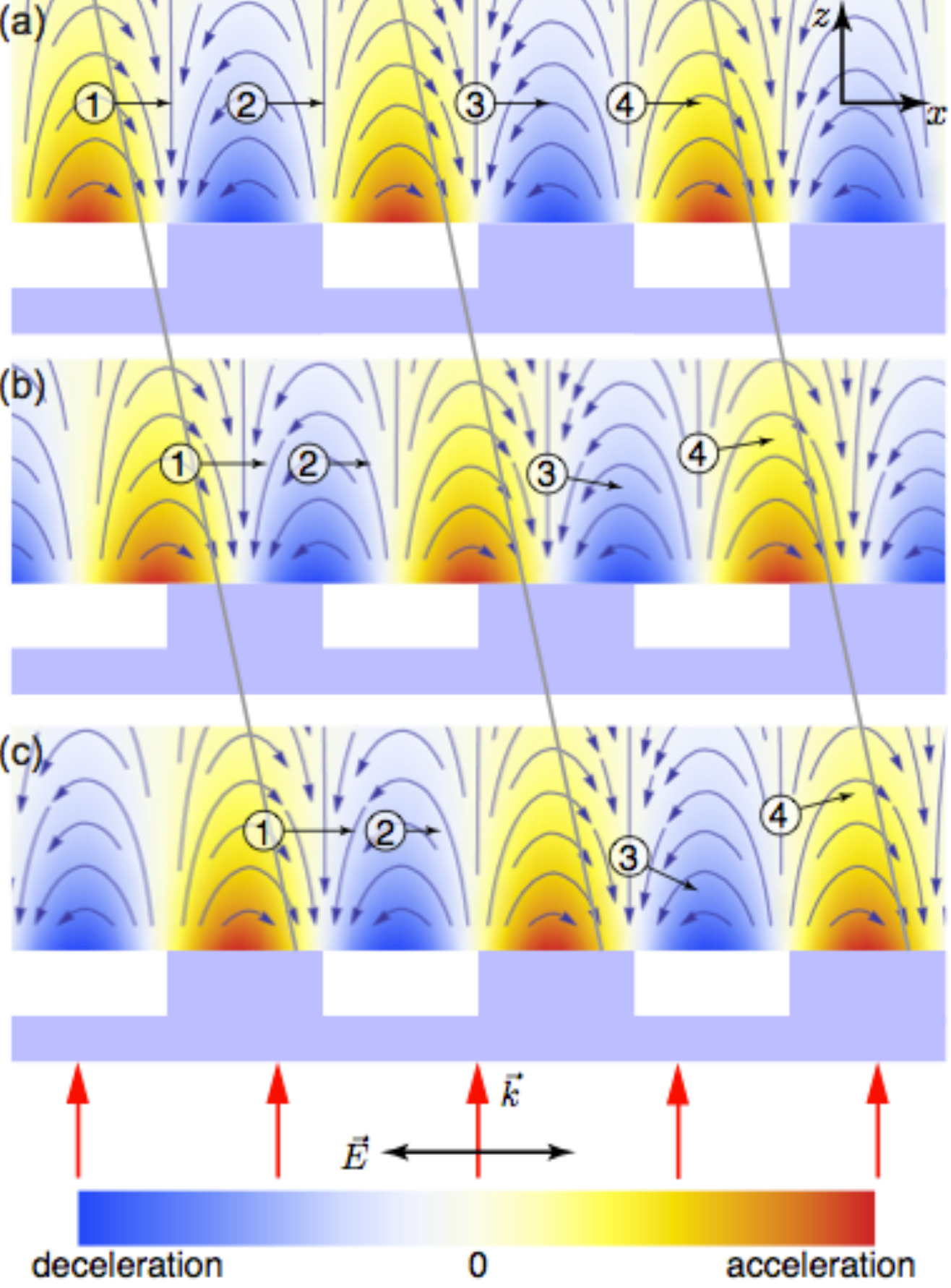}
	\caption{(a)-(c), three subsequent conceptual pictures of four charged particles (circles) passing the transparent grating (light blue structure), which is illuminated by the laser from below, polarized in the plane of projection. Time step between each picture: 1/4 optical period. The electric field of the first spatial harmonic, which is synchronous with the electrons, falls off exponentially away from the grating (color-coded). The charged particles are assumed to be positrons. Depending on the relative position of the positron inside the laser field the force onto the positron can be accelerating (1), decelerating (2) or deflecting (3,4). Note that the geometry implies that the fields are transverse magnetic (TM), so the only field components are $E_x$, $E_z$ and $B_y$ ($\varphi=\psi=\theta=0$ in Fig.\ \ref{fig:coord_sys}). }
	\label{fig:figure1}
\end{figure}

Fig.\ \ref{fig:figure1} shows the concept of synchronous particle acceleration exploiting the first spatial harmonic of a transparent grating, which is excited by a laser beam incident perpendicularly to the grating surface.

It is important to note that the longitudinal force always goes along with a transverse component that causes deflection of the particles. However, the accelerating and deflecting forces are out of phase, which implies that stable acceleration, i.e., temporal bunching leads to a defocusing of the electron bunch. This can be overcome either by alternating phase focusing \cite{Good1953,Swenson1976,Wangler2008} or by a biharmonic structure that focuses the bunch by ponderomotive interaction with asynchronous modes \cite{Naranjo2012}.

The issue of a skew acceleration pattern remains, namely the exponential dependence of the accelerating force as a function of the particle's distance from the surface. It has however been shown that by using two parallel surfaces and creating phase-stabilized fields on both of them, one can arrange the setup in such a manner that the deflecting forces cancel each other on the axis of the accelerator, creating a symmetric force pattern \cite{plettner2009,Plettner2011}. The distance between the two surfaces has to be on the order of or ideally smaller than $\delta$ in order to efficiently accelerate particles at the center of such a double grating structure.

Equation \ref{eq:int8} also reveals that a single grating cannot be used to accelerate speed-of-light particles ($\beta\sim1$) unless $\varphi\neq0$ \cite{palmer1980}. In contrast, for a double grating structure exists a speed-of-light mode, which can be used to continuously accelerate $\beta\sim1$ particles. It arises when $k_{\|}=k_0$ and hence $d^2\mathbf{A}/dz^2 =0$ according to Eq.\ \ref{eq:int2}. It can be solved by $\mathbf{A}(\mathbf{r})=\mathbf{A}(x,y) (1+\kappa z)$. In the case of a single surface this mode cannot exist because a constant or linearly increasing electric field extending to infinity is unphysical, but in the presence of a second boundary the linear solution is meaningful implying a constant longitudinal (accelerating) force component \cite{Plettner2011}.

In the following we will restrict our discussion to a TM mode with $\varphi=\psi=\theta=0$, as shown in Fig.\ \ref{fig:figure1}. Hence,
\begin{equation}
\mathbf{F_{r}}=q\left(
\begin{array}{c}
icB_y /(\beta\gamma)\\
0\\
-cB_y/(\beta\gamma^2)
\end{array}\right).
\label{eq:int9}
\end{equation}

%In order to use laser electric fields, it was first proposed to use a single metallic grating in combination with a static magnetic field in order to cancel the deflection forces \cite{palmer1980}. Here, it has to be $\varphi\neq 0$ in order to accelerate speed-of-light particles. It has also been suggested that suitable fields can be generated close to a photonic dielectric grating structure \cite{plettner2006}.

\section{Dephasing of an accelerating particle}
In order to satisfy the synchronicity condition for an accelerating non-relativistic particle, the phase velocity of the accelerating mode has to change continuously to account for the change in velocity of the particle. In case of a constant phase velocity of the accelerating mode the relative phase of the particle with respect to the mode changes as the particle gains speed. We call this effect \textit{dephasing}. We now derive an estimate of the distance over which acceleration can take place if the mode remains at the same phase velocity, i.e.: When does the particle become accelerated so much that it starts experiencing deceleration upon dephasing?

The only assumption is $\Delta\beta/\beta \ll 1$ during the acceleration. This is true for relativistic $\beta\sim 1$ particles as the change in velocity in the laboratory frame is practically zero. In the non-relativistic case this assumption is valid as long as the particle's energy gain $\Delta E$ over one wavelength $\lambda$ of the driving field is well below the particle's rest energy $m_0 c^2$, i.e., $G\ll m_0 c^2/\lambda$. We note that the relativistic factor $a_{0}$ (=$\frac{q E_{0}}{mc^{2}k_{0}}$) is typically $10^{-3}$ or smaller in the context of DLAs, so that this assumption is valid. In contrast, for larger acceleration gradients $G> m_0 c^2/\lambda$, the particles can be accelerated from rest to relativistic energies within one cycle of the driving field.

The accelerating force can be written as
\begin{equation}
F_x(x,z,t)=G(z)\text{Re}\left[e^{i(k_0x/\beta_0-\omega t + \phi)}  \right],
\label{eq:deph1}
\end{equation}
with $G(z)=G \exp\left(-k_0z/(\beta\gamma)\right)$. In the electron's co-moving frame $\omega t=\omega x/(\beta(x)c)=k_0x/\beta(x)$, with the instantaneous velocity $\beta(x)=\beta_0+\Delta\beta(x)$. We derive the instantaneous acceleration gradient
\begin{equation}
\tilde{G}(x,z)=F_x(x,z)=G(z)\cos\left(\frac{k_0x}{\beta_0^2}\int_0^x{\beta'(u)du}+\phi\right),
\label{eq:deph5}
\end{equation}
using
\begin{equation}
\begin{split}
\frac{\beta_0}{\beta(x)}&=\frac{\beta_0}{\beta_0+\Delta\beta(x)}\\
&=\frac{1}{1+\frac{\int_0^x\beta'(u)du}{\beta_0}}\approx 1-\frac{\int_0^x\beta'(u)du}{\beta_0},
\end{split}
\label{eq:deph4}
\end{equation}
with $\beta'=d\beta/dx$. The validity of Eq.\ \ref{eq:deph5} can be verified with the simulations presented below (Fig.\ \ref{fig:figure5}). The first term in the cosine is the dephasing term, which we estimate here. With the kinetic energy $E_{\mathrm{kin}}=m_0c^2(\gamma-1)$, we derive
\begin{equation}
\frac{d\beta}{dx}=\frac{d}{dx}\sqrt{1-\frac{1}{\gamma^2}}=\frac{1}{\gamma^3\sqrt{1-\frac{1}{\gamma^2}}}\frac{d\gamma}{dx}=\frac{G}{m_0c^2\beta\gamma^3},
\end{equation}
with the acceleration gradient $G=dE_\mathrm{kin}/dx$. Hence, the dephasing angle for a given distance $z_0$ from the grating surface
\begin{equation}
\begin{split}
\Delta\phi(x)&:=\frac{k_0x}{\beta_0^2}\int_0^x{\beta'(u)du}\\
&=\frac{k_0x}{m_0 c^2\beta_0^2}\int_0^x{\frac{\tilde{G}(u,z_0)}{\beta(u)\gamma^3(u)}du} <\frac{k_0x^2G_{\mathrm{max}}}{m_0 c^2\beta_0^3\gamma_0^3}.
\end{split}
\label{eq:deph9}
\end{equation}
This inequality holds because for an \textit{accelerating} particle $\tilde{G}(u,z_0)<G_{\mathrm{max}}$, the maximum acceleration gradient, $\beta(u)>\beta_0$, the initial velocity, and $\gamma(u)>\gamma_0=(1-\beta_0^2)^{-1/2}$. Using $\beta^2\gamma^2=\gamma^2-1=(\gamma-1)(\gamma+1)$ and demanding that the dephasing has to be smaller than $\pi/2$ for the acceleration to take place, we obtain the estimate for the maximum length, over which a particle can be accelerated until acceleration just ceases. This \textit{dephasing length} is
\begin{equation}
x_{\mathrm{deph}}=\left(\frac{\beta_0\lambda_0 E_{\mathrm{kin}} \left(\frac{E_{\mathrm{kin}}}{m_0c^2}+1\right)\left(\frac{E_{\mathrm{kin}}}{m_0c^2}+2\right)}{4 G_{\mathrm{max}}}\right)^{1/2}.
\label{eq:deph11}
\end{equation}
This approximation includes the intuitively right behavior: relativistic particles dephase after a longer distance, and a larger acceleration gradient causes dephasing to set in more quickly. We calculate the dephasing lengths for non-relativistic and relativistic electrons inside a double grating structure below (Table\ \ref{tab:dephasing} on page \pageref{tab:dephasing}).

\section{Simulation of acceleration at a single dielectric grating}

There is a variety of methods to simulate the propagation of electromagnetic waves through media, for example, the finite-difference time-domain (FDTD) method \cite{yee1966}, the finite-element method (FEM) \cite{Coccioli1996,Ferrari2007}, the finite integration technique (FIT) \cite{Weiland1977} or the pseudospectral time domain (PSTD) method \cite{Liu2005}. We chose yet another method, namely an eigenmode expansion method \cite{Pai1991} for our calculations of a laser pulse propagating through a dielectric grating. It is used to compute the amplitudes of the spatial harmonics of a grating with an infinitely periodic, rectangular profile. The method can be used to directly calculate the amplitudes $E_y$ and $B_y$ of the TE and TM mode in Eq.\ \ref{eq:int6}. We verified this method by comparison to published results \cite{Tremain1978,Pai1991}.

\begin{figure*}
	\centering
		\includegraphics[width=\textwidth]{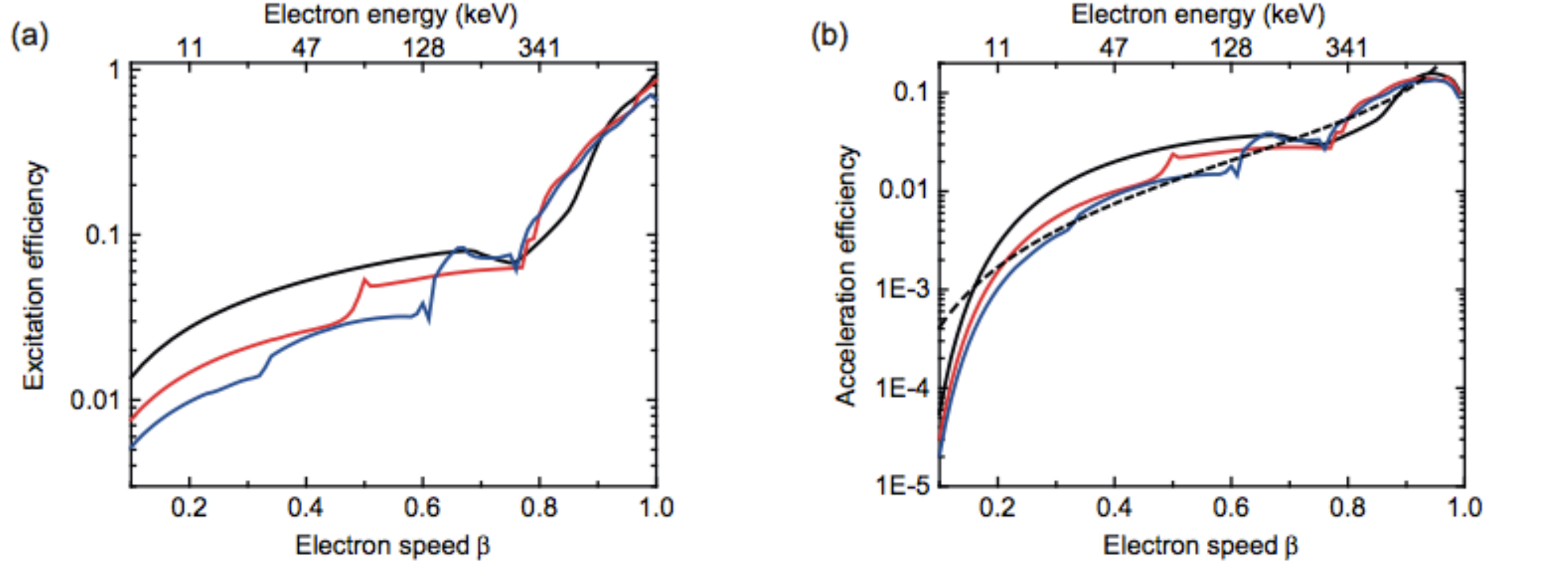}
	\caption{(a), Excitation efficiency $\epsilon_\mathrm{exc}=cB_y^{(n)}/E_\mathrm{p}$ of the $n$-th spatial harmonic (first: black, second: red, third: blue) as a function of the electron velocity $\beta$ (bottom axis) and of the electron energy $E_\mathrm{kin}$ (top axis). The grating period is $\lambda_\mathrm{p}(\beta)=n\beta\lambda$. The exciting laser wavelength is $\lambda=800$\,nm. The aspect ratio (i.e., the ratio of the trench width to the grating period) and grating depth have been optimized to maximize $cB_y^{(n)}$ for each $\beta$. During the optimization we first varied the aspect ratio between 0 and 1 for a fixed grating depth of 250\,nm. Afterwards we varied the grating depth between 0 and $\lambda_\mathrm{p}$ for the optimum aspect ratio. (b), The acceleration efficiency $\epsilon_\mathrm{acc}=G(z_0)/(eE_\mathrm{p})$ at a distance of $z_0=100$\,nm from the grating surface as a function of $\beta$ (bottom axis) and $E_\mathrm{kin}$ (top axis), exploiting the first (black), second (red) and third (blue) spatial harmonics. For highly relativistic velocities ($\beta\sim 1$) $\epsilon_\mathrm{acc}$ drops to zero with $\sqrt{1-\beta^2}$ for a single grating (Eq.\ \ref{eq:int9}). The black dashed line shows the linear fit of $\epsilon_\mathrm{exc}=E_\mathrm{kin}\cdot1.61\cdot 10^{-7}\,\mathrm{eV}^{-1}$  up to an energy of 1\,MeV for the first spatial harmonic. The kinks in the efficiency in (a) and (b), e.g., for the second spatial harmonic at $\beta\sim1/2$ and for the third spatial harmonic at $\beta\sim2/3$, occur when the next lower order spatial harmonic starts to propagate, i.e., $k_0>k^{(n-1)}=(n-1)k_\mathrm{p}=(n-1)k_0/(n\beta)$ and hence $\beta>(n-1)/n$.}
	\label{fig:figure3}
\end{figure*}

Again we focus on the TM case as shown in Fig.\ \ref{fig:figure1} because of the longitudinal accelerating electric field component. In the simulation we choose the exciting laser wavelength and determine the grating period such that the $n$-th spatial harmonic is synchronous with electrons with velocity $\beta c$. Hence, the grating period is given by
\begin{equation}
\lambda_\mathrm{p}=n\beta\lambda.
\label{eq:sim0}
\end{equation}
We directly simulate the amplitude $B_y^{(n)}$ of the $n$-th harmonic and therefore obtain the acceleration efficiency
\begin{equation}
\epsilon_\mathrm{acc}:=\frac{G(z_0)}{eE_\mathrm{p}}.
\label{eq:sim1}
\end{equation}
Here $G(z_0)$ is the acceleration gradient
\begin{equation}
G(z_0)=\frac{c\left|B_y^{(n)}\right|}{\beta\gamma}\exp\left(-\frac{k_0z_0}{\beta\gamma}\right),
\label{eq:sim2}
\end{equation}
at a fixed distance $z_0$ from the grating surface, $E_p$ is the exciting laser peak electric field and $e$ is the elementary charge. We optimize the grating depth and aspect ratio to maximize $B_y^{(n)}$.

In Fig.\ \ref{fig:figure3} we show the excitation efficiency $\epsilon_\mathrm{exc}:=cB_y^{(n)}/E_\mathrm{p}$ and the acceleration efficiency $\epsilon_\mathrm{acc}$ for the first, second and third spatial harmonic as a function of the electron speed $\beta$, which directly determines the grating period (Eq.\ \ref{eq:sim0}). It becomes clear that using higher order spatial harmonics is less efficient than using the fundamental, which can be understood from diffraction effects. It can further be seen that the efficiency to excite a given spatial harmonic increases as $\beta\rightarrow 1$, which is due to wave matching between the wave vector of the incident laser $k_0$ and that of the synchronous spatial harmonic $k=k_0/\beta$. For highly relativistic velocities ($\beta\sim1$) $\epsilon_\mathrm{acc}$ vanishes in the case of a single grating as discussed above, but can be on the order of one for double grating structures \cite{plettner2006}. We show a linear fit of $\epsilon_\mathrm{acc}$, which allows us to estimate the length of a non-relativistic DLA below.

To gain further insight we perform a particle tracking in the resulting fields above the grating including all spatial harmonics. This way deflecting forces and dephasing effects can be studied. We assume a single electron passing the grating surface with a velocity $\beta c$ and an exciting laser pulse, incident perpendicularly to the electron's trajectory, with an optical electric field $\tilde{E}_\mathrm{p}\propto\exp\left(-(x/w_\mathrm{l})^2-2\text{ln}(2)(t/\tau_\mathrm{p})^2\right)$, with $w_\mathrm{l}$ the 1/$e$ focal waist radius and $\tau_\mathrm{p}$ the pulse duration (full width at half maximum of the intensity envelope). Hence, in its co-moving frame the electron experiences the instantaneous electric field $E_\mathrm{p}\exp\left(-(x/w_\mathrm{int})^2\right)$ with the characteristic interaction distance
\begin{equation}
w_\mathrm{int}=\left(\frac{1}{w_\mathrm{l}^2}+\frac{2\text{ln}(2)}{(\beta c \tau_\mathrm{p})^2}\right)^{-1/2}.
\label{eq:sim3}
\end{equation}

\begin{figure*}
	\centering
		\includegraphics[width=\textwidth]{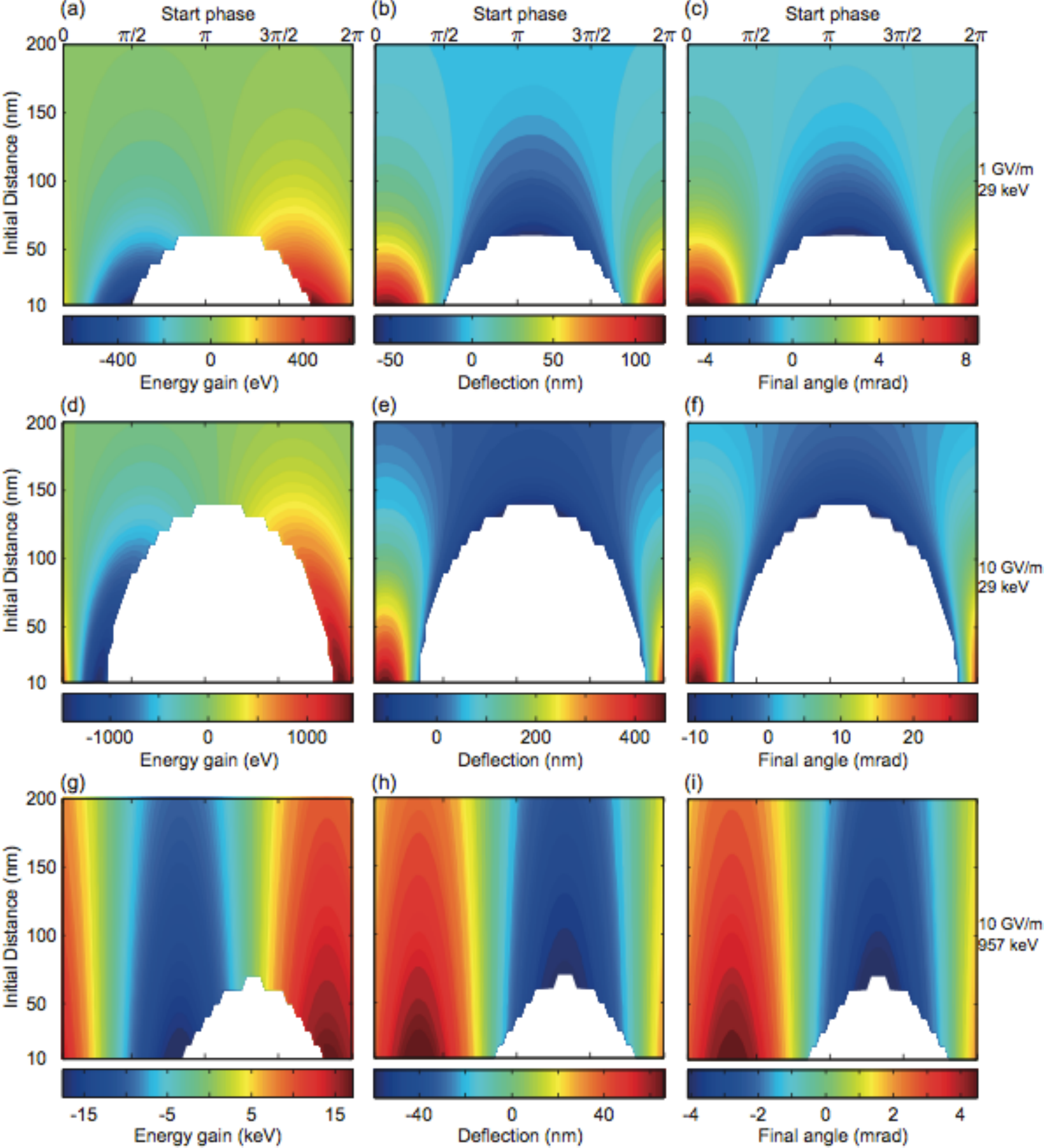}
	\caption{Particle tracking results of a single electron interacting with laser pulses in close proximity of a fused silica grating. The laser parameters are: wavelength $\lambda=800$\,nm, focal waist radius $w_\mathrm{l}=5$\,$\mu$m, pulse duration $\tau_\mathrm{p}=100$\,fs, laser peak electric field $E_\mathrm{p}=1$\,GV/m (a-c) and $E_\mathrm{p}=10$\,GV/m (d-i). The initial electron energy is $E_\mathrm{kin}=29$\,keV ($\beta=0.33$) (a-f) and $E_\mathrm{kin}=957$\,keV ($\beta=0.94$) (g-i). The first spatial harmonic interacts synchronously with the electrons, hence $\lambda_\mathrm{p}=260$\,nm (a-f) and $\lambda_\mathrm{p}=750$\,nm (g-i). Color-coded plots show the energy gain $\Delta E_\mathrm{kin}$ (a,d,g), the deflection $\Delta z$ (b,e,h) and the final angle $\beta_z/\beta_x$ (c,f,i) as a function of the initial distance from the grating, i.e.\ the distance of the electron from the grating surface before the interaction with the laser, and of the relative start phase between the electron and the laser field. The transverse decay length of the acceleration is $\delta=45$\,nm (a-f) and $\delta=350$\,nm (g-i). For the white areas the electron becomes deflected into the grating during the simulation. The characteristic interaction distance is $w_\mathrm{int}=4.3$\,$\mu$m (a-f) and $w_\mathrm{int}=4.9$\,$\mu$m (g-i).}
	\label{fig:figure4}
\end{figure*}

In Fig.\ \ref{fig:figure4} we show the results of the particle tracking for non-relativistic and relativistic electrons interacting with the first spatial harmonic close to a fused silica grating. The maximum laser peak electric field of $E_\mathrm{p}=10$\,GV/m corresponds to a peak fluence $F=1.42$\,J/cm$^2$, close to the damage threshold for a 110 fs pulse length \cite{Soong2011b}. The dependencies of the accelerating fields, derived above, can be clearly seen, for example, the phase shift between the accelerating and deflecting force (Eq.\ \ref{eq:int9}) or the larger decay distance $\delta$ for relativistic electrons (Fig.\ \ref{fig:figure4} (g-i)) as compared to non-relativistic electrons (Fig.\ \ref{fig:figure4} (a-f)). Note that in our simulation a larger start phase corresponds to a later start time. Hence, bunching takes place when electrons with a smaller start phase (earlier start time) become less accelerated than those with a larger start phase (later start time). The strong deflection of the non-relativistic electrons suggests choosing laser peak electric fields well below 10\,GV/m for the acceleration of non-relativistic electrons in order to prevent beam loss as well as surface charging of the dielectric material that can cause further deflection. Of course this is not necessary if a microbunched electron beam, phase-stabilized to the laser field, with a microbunch duration much smaller than an optical cycle is used and if the electrons only occupy start phases for which deflection is small.

\begin{figure*}
	\centering
		\includegraphics[width=\textwidth]{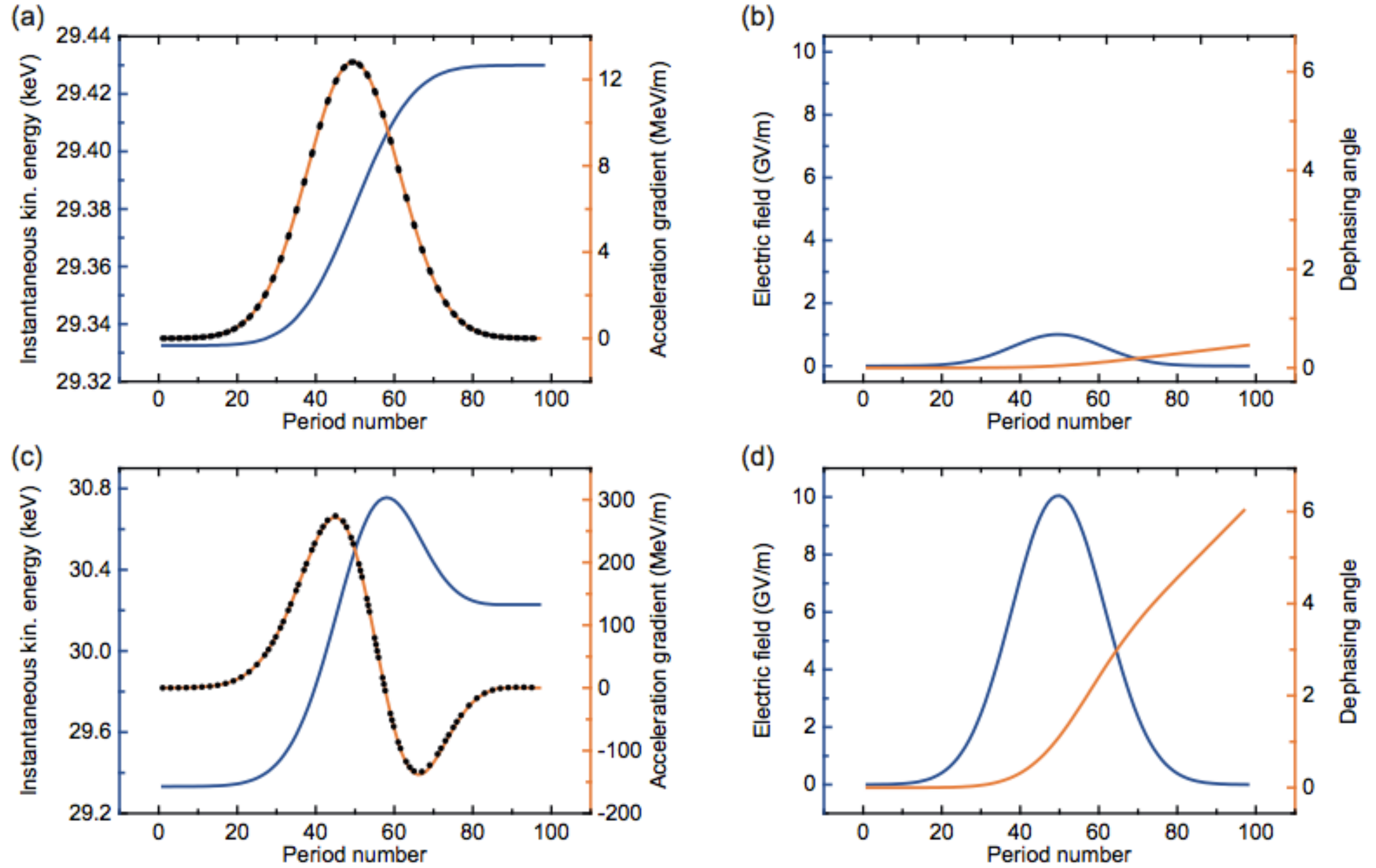}
	\caption{We show the instantaneous parameters of a single electron interacting with laser pulses in close proximity of a fused silica grating as a function of the number of grating periods passed ($x$-coordinate). We include the instantaneous kinetic energy $\tilde{E}_\mathrm{kin}(x)$ ((a,c), blue curve), acceleration gradient $\tilde{G}(x)$ ((a,c), orange curve, black dots), laser electric field $\tilde{E}_\mathrm{p}(x)$ ((b,d), blue curve) and dephasing angle $\Delta\phi(x)$ ((b,d), orange curve). The acceleration gradient has been directly derived from $\tilde{E}_\mathrm{kin}(x)$ via $\tilde{G}(x)=\mathrm{d}\tilde{E}_\mathrm{kin}(x)/\mathrm{d}x$ (orange curve) and fitted according to Eq.\ \ref{eq:deph5} with $G(x,z)=\epsilon \tilde{E}_\mathrm{p}(x) \exp\left(-z/\delta\right) \cos\left(\Delta\phi(x)\right)$ (black dots) with the free fit parameter $\epsilon$. (a,b), Identical simulation parameters as used in Fig.\ \ref{fig:figure4}\,(a-c) for a start phase of 1.6\,$\pi$ and an initial distance $z_0=100$\,nm. For those parameters dephasing and deflection can be neglected and therefore the fit parameter $\epsilon=0.013$, which relates the acceleration gradient to the applied laser peak electric field, equals $\epsilon_\mathrm{acc}$ (Fig.\ \ref{fig:figure3}\,(b)) for the electron speed $\beta=0.33$. (c,d), Identical simulation parameters as used in Fig.\ \ref{fig:figure4}\,(d-f) for a start phase of 1.8\,$\pi$ and an initial distance $z_0=50$\,nm. Here, dephasing is so severe that after initial acceleration the electron becomes decelerated after passing $\sim$60 grating periods.}
	\label{fig:figure5}
\end{figure*}

In Fig.\ \ref{fig:figure5} we show the instantaneous kinetic energy $\tilde{E}_\mathrm{kin}(x)$, the instantaneous acceleration gradient $\tilde{G}(x)$, the laser electric field $\tilde{E}_\mathrm{p}(x)$ and the dephasing angle $\Delta\phi(x)$ as a function of the longitudinal position of the electron. In these simulation results, the width of the instantaneous laser electric field $\tilde{E}_\mathrm{p}(x)$ equals the characteristic interaction distance $w_\mathrm{int}=4.3$\,$\mu$m. The connection between the laser electric field, the dephasing angle and the instantaneous acceleration gradient confirms Eq.\ \ref{eq:deph5}. In Fig.\ \ref{fig:figure5}\,(c,d) the dephasing of the electron is so severe that it experiences deceleration after initial acceleration.

\section{Simulation of acceleration at a double grating structure}

We call the geometry with two gratings facing each other the double grating structure. It exhibits the advantage of enabling a symmetric acceleration pattern because the synchronous mode of a double grating structure does not decay exponentially with increasing distance from the grating surface, as it is the case for a single grating. Instead the field pattern at a distance $z$ is given by
\begin{equation}
B_y=(C_\mathrm{s}\sinh(k_z z) + C_\mathrm{c} \cosh(k_z z)) \cos(k_x x-\omega t),
\label{eq:double0}
\end{equation}
with $k_x=k_0/\beta$, $k_z=k_0/(\beta\gamma)$ (Eq.\ \ref{eq:int4c}), and $C_\mathrm{s}$ and $C_\mathrm{c}$ constants \cite{plettner2009,Plettner2011}. Double gratings also support a mode that can travel synchronously with $\beta=1$ particles \cite{Plettner2011}, as mentioned above. Following the same discussion as around Eq.\ \ref{eq:int9}, we obtain for the force vector
\begin{widetext}
\begin{equation}
\mathbf{F_{r}}=qc\left(
\begin{array}{c}
\frac{1}{\beta\gamma}\left(C_\mathrm{s}\cosh(k_z z)+C_\mathrm{c}\sinh(k_z z)\right)\sin(k_x x-\omega t)\\
0\\
-\frac{1}{\beta\gamma^2}\left(C_\mathrm{s}\sinh(k_z z)+C_\mathrm{c}\cosh(k_z z)\right)\cos(k_x x-\omega t)
\end{array}\right).
\label{eq:double1}
\end{equation}
\end{widetext}

\begin{figure*}
	\centering
		\includegraphics[width=\textwidth]{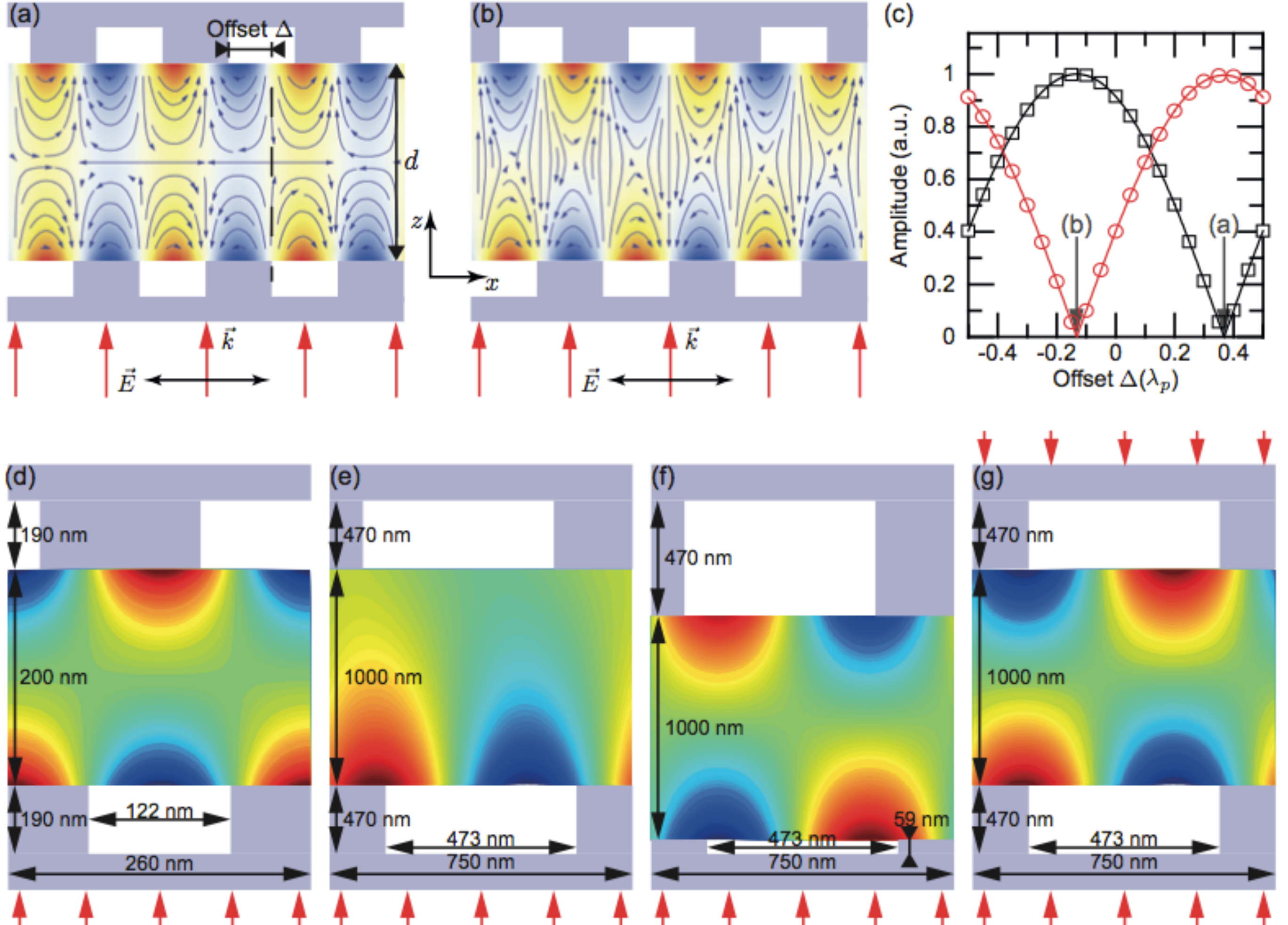}
	\caption{(a,b), Conceptual pictures of the dielectric double grating structure (light blue), which is illuminated by the laser from below. The separation between the two grating surfaces is $d$ and the upper grating has a longitudinal offset $\Delta$ compared to the lower one. The accelerating electric field profile of the first spatial harmonic (color-coded, red: acceleration, blue: deceleration) as a function of $z$ can be $\cosh(k_z z)$ (a), $\sinh(k_z z)$ (b) or a superposition thereof. (c), Simulated relative amplitudes $C_\mathrm{c}$ (black squares) and $C_\mathrm{s}$ (red circles) as a function of the offset $\Delta$ between upper and lower grating. The solid lines are fitted curves with the fit function $\left|\sin(\pi\Delta)\right|$. The grating parameters are: grating period $\lambda_\mathrm{p}=260$\,nm, grating depth: 190\,nm, trench width: 0.47\,$\lambda_\mathrm{p}$, optimized for a maximum excitation efficiency of the single grating, and grating distance $d=200$\,nm. The laser wavelength is $\lambda=800$\,nm. We indicate the offset parameter used in (a,b) by grey arrows. (d-g), Here we show the \textit{magnetic field} profile $B_y$ of the first spatial harmonic for four different double grating geometries (grating parameters are shown in the sketch). In (d) the offset between the upper and lower grating is $\Delta=0.37\,\lambda_\mathrm{p}$. We obtain $C_\mathrm{s}=9.2\cdot 10^{-3} E_\mathrm{p}/c$. %The value $cC_\mathrm{s}/(\beta\gamma E_\mathrm{p})=2.7\cdot 10^{-2}$ can be directly compared to $\epsilon_\mathrm{acc}(\beta=0.33)=1.3\cdot 10^{-2}$ in Fig.\ \ref{fig:figure3}(b).
(e), The grating parameters of the lower and upper grating correspond to the optimized values used in Fig.\ \ref{fig:figure4}\,(g-i). The offset is $\Delta=-0.08\,\lambda_\mathrm{p}$. The excitation efficiency of the spatial harmonic at the upper grating is negligible. Hence, the field profile looks (almost) like in the case of a single grating. (f), The grating depth of the lower grating has been decreased to reduce the excitation efficiency at the lower grating compared to the upper one, which leads to the desired $\sinh$-profile for one-sided illumination ($C_\mathrm{s}=0.05 E_\mathrm{p}/c$). (g), Symmetric illumination also yields the favorable $\sinh$-profile with the advantage of maximum efficiency ($C_\mathrm{s}=0.25 E_\mathrm{p}/c$). Color code: linear from blue to red; green: 0.}
	\label{fig:figure6}
\end{figure*}

In Fig.\ \ref{fig:figure6}\,(a) we show the electric field profile of a double grating structure, for which $C_\mathrm{s}=1$ and $C_\mathrm{c}=0$. As for the single grating, regions of acceleration and deceleration exist. However, around the axis of the structure the accelerating force component is rather uniform, as can be seen from $dF_x/dz\propto d\cosh(k_z z)/dz|_{z=0}=0$. Moreover, the transverse force component can focus electrons towards the axis. %, because $F_z(z<0)\propto -\sinh(k_z z)|_{z<0}>0$ and $F_z(z>0)\propto -\sinh(k_z z)|_{z>0}<0$.
The longitudinal and transverse forces are out of phase. Therefore an electron passing through the structure at a relative position to the field where it is maximally accelerated does not experience any focusing force, and vice versa. In Fig.\ \ref{fig:figure6}\,(b), where $C_\mathrm{c}=\mathrm{max}$ and $C_\mathrm{s}=0$, the accelerating force component $F_x$ vanishes on axis, as $\sinh(k_z z)|_{z=0}=0$. Hence, this field profile is not suitable for particle acceleration.

To obtain the field distributions shown in Fig.\ \ref{fig:figure6}\,(a,b) we again use the eigenmode expansion method \cite{Pai1991} \footnote{Note that the speed-of-light mode, which only occurs for $\lambda_\mathrm{p}=\lambda$, is not included in this simulation. However, it can be included as shown in \cite{Plettner2011}. Further simulations are needed to investigate how a potentially excited speed-of-light mode affects the acceleration gradient of 1\,MeV electrons.}. We calculate the amplitudes of the spatial harmonics excited by a single laser source, impinging from below. In Fig.\ \ref{fig:figure6}\,(c) we show the simulated relative amplitudes $C_\mathrm{c}$ and $C_\mathrm{s}$ as a function of the relative longitudinal offset $\Delta$ between the upper and the lower grating, i.e., the shift between the grating grooves of the upper grating as compared to the lower one. The offset can be related to a time delay between the excitation of the two single gratings. By changing the offset the relative phase between the exponentially decaying fields at both grating surfaces is shifted and hence a $\cosh$- or $\sinh$-profile can be realized.

For a dielectric laser accelerator only the $\sinh$-profile of the magnetic field that implies a $\cosh$-profile of the accelerating force ($C_\mathrm{c}=0$ and $C_\mathrm{s}=\mathrm{max}$ in Fig.\ \ref{fig:figure6}\,(d)) leads to useful acceleration \cite{plettner2009,Plettner2011}. In Fig.\ \ref{fig:figure6}\,(e) the excitation of the spatial harmonic at the lower grating is more efficient than at the upper one and therefore the center of the $\sinh$-profile does not coincide with the center of the vacuum channel. The difference in excitation efficiency between the upper and lower grating can be equalized by changing the grating parameters of the lower grating (Fig.\ \ref{fig:figure6}\,(f)) at the cost of a lower acceleration efficiency.

Alternatively, a symmetric mode profile can be achieved by pumping the double grating structure from both sides (Fig.\ \ref{fig:figure6}\,(g)) \cite{plettner2009,Aimidula2014} with the advantage of maximum efficiency. Note that for symmetric illumination the offset has to be either zero or half the grating period, because in any other case the 1$^{\mathrm{st}}$ and -1$^{\mathrm{st}}$ spatial harmonic of the two single gratings are not excited equally strong. This would lead to a skew acceleration profile, because for the upward propagating laser beam the 1$^{\mathrm{st}}$, and for the downward propagating laser beam the -1$^{\mathrm{st}}$ spatial harmonic is synchronous with electrons passing through the structure from left to right.

\begin{figure*}
	\centering
		 \includegraphics[width=\textwidth]{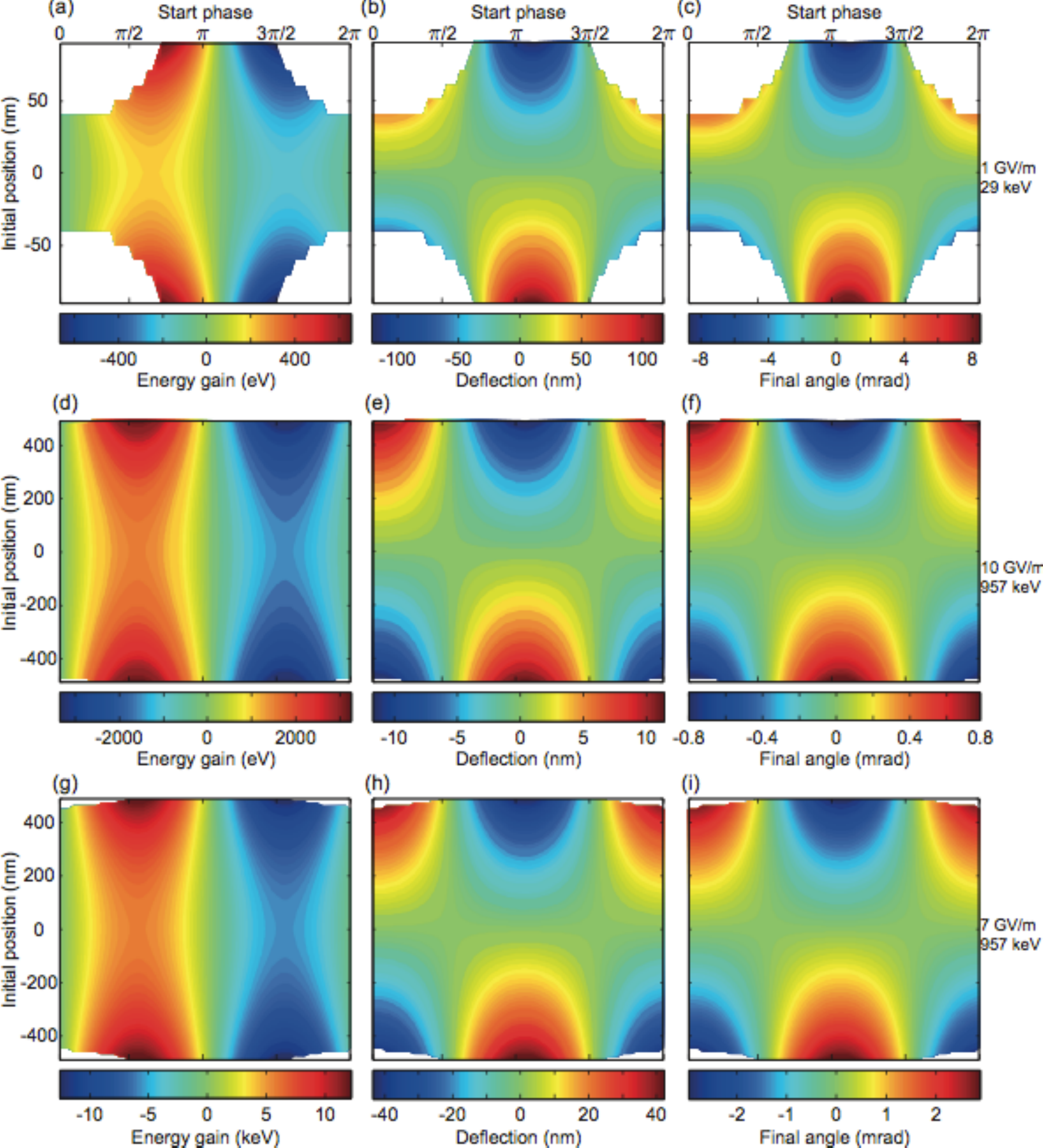}
	\caption{Particle tracking results of a single electron interacting with laser pulses in a fused silica double grating structure. Here we neglect all asynchronous modes. All white areas correspond to injection phases and transverse positions at which electrons crashed into the grating structures. The laser parameters are: wavelength $\lambda=800$\,nm, focal waist radius $w_\mathrm{l}=5$\,$\mu$m, pulse duration $\tau_\mathrm{p}=100$\,fs, laser peak electric field $E_\mathrm{p}=1$\,GV/m (a-c), $E_\mathrm{p}=10$\,GV/m (d-f) and $E_\mathrm{p}=7$\,GV/m (g-i). The initial electron energy is $E_\mathrm{kin}=29$\,keV ($\beta=0.33$) (a-c) and $E_\mathrm{kin}=957$\,keV ($\beta=0.94$) (d-i). (a-c), Grating parameters from Fig.\ \ref{fig:figure6}\,(d), single sided pumping. (d-f), Grating parameters from Fig.\ \ref{fig:figure6}\,(f), also single-sided pumping. (g-i), Grating parameters from Fig.\ \ref{fig:figure6}\,(g), double-sided pumping. Color-coded plots show the energy gain $\Delta E_\mathrm{kin}$ (a,d,g), the deflection $\Delta z$ (b,e,h) and the final angle $\beta_z/\beta_x$ (c,f,i) as a function of the initial offset from the vacuum channel axis and of the relative start phase between the electron and the laser field. For the white areas the electron becomes deflected into the grating during the simulation. The characteristic interaction distance is $w_\mathrm{int}=4.3$\,$\mu$m (a-c) and $w_\mathrm{int}=4.9$\,$\mu$m (d-i).}
	\label{fig:figure7}
\end{figure*}

Particle tracking results for relativistic and non-rel\-a\-tiv\-istic electrons in a double grating structure are shown in Fig.\ \ref{fig:figure7}. In this simulation we consider the geometries depicted in Fig.\ \ref{fig:figure6}\,(d,f,g). We simulate the energy gain, the deflection, as well as the final angle of the electron trajectory with respect to the grating surface. The results can be directly compared with the single grating simulation in Fig.\ \ref{fig:figure4}. The advantage of the symmetric field pattern can clearly be seen: the energy gain for electrons passing the grating at a distance $z_0$ from the axis equals the gain for electrons passing at $-z_0$. Moreover there exists an initial start phase ($\sim\pi$) for which electrons starting at positive $z_0$ are deflected downwards and electrons starting at negative $z_0$ are deflected upwards. This corresponds to a focusing force towards the axis of the structure. It can also be seen that the acceleration and focusing are out of phase ($\pi/2$ phase shifted), which is expected from Eq.\ \ref{eq:double1}. %Note that in the case of symmetric illumination by two lasers the peak electric field has to be reduced by a factor of $1/\sqrt{2}$ to remain the same fluence and prevent damage to the glass.

In Fig.\ \ref{fig:figure7}\,(a-c) we show the particle tracking results of non-relativistic 30\,keV electrons passing through the double grating structure depicted in Fig.\ \ref{fig:figure6}\,(d) for a laser peak electric field of $E_\mathrm{p}=1$\,GV/m. The maximum on-axis energy gain of $\Delta E=200$\,eV corresponds to a maximum acceleration gradient of $G_\mathrm{max}=\Delta E/(\sqrt{\pi}w_\mathrm{int})=26$\,MeV/m, using $w_\mathrm{int}=4.3$\,$\mu$m. Note that if a continuous electron beam is passing through the structure, the beam diameter should be smaller than $\sim$100\,nm in order to prevent electrons from being deflected into the grating. This strongly suggests using microbunched electron beams with a sub-laser-cycle microbunch duration, that are injected at start phases for which the structure acts both accelerating and focusing, i.e., between $\pi/2$ and $\pi$.

We further perform particle tracking simulations of relativistic 1\,MeV electrons inside double grating structures. In Fig.\ \ref{fig:figure7}\,(d-f), corresponding to the geometry shown in Fig.\ \ref{fig:figure6}\,(f), we use a laser peak field of $E_\mathrm{p}=10$\,GV/m. The maximum on-axis energy gain is $\Delta E=1.5$\,keV and translates into $G_\mathrm{max}=\Delta E/(\sqrt{\pi}w_\mathrm{int})=170$\,MeV/m ($w_\mathrm{int}=4.9$\,$\mu$m).

In Fig.\ \ref{fig:figure7}\,(g-i) 1\,MeV electrons pass through the symmetrically pumped structure shown in Fig.\ \ref{fig:figure6}\,(g). Here we choose a laser peak electric field of $E_\mathrm{p}=7$\,GV/m in each laser beam, reduced by a factor of $1/\sqrt{2}$ compared to the single beam illumination. Therefore the overall fluence remains constant and damage to the grating is prevented. Hence, we can directly compare the two-beam with the single-beam excitation. The maximum on-axis energy gain is $\sim$5.7\,keV, translating into $G_\mathrm{max}\simeq 650$\,MeV/m.

With a narrower, 300\,nm wide vacuum channel we expect from Eq.\ \ref{eq:double1} the gradient to be about twice as large, i.e.\ 1.3\,GeV/m. With slightly different parameters and/or grating materials, the gradient could exceed 10\,GeV/m \cite{plettner2006}. Note that \cite{plettner2006} considers the speed-of-light mode only.

From these simulations it becomes clear that relativistic electrons are more efficiently accelerated and less affected by deflecting forces. Together with the other advantage of wider possible vacuum channels in double grating structures designed for relativistic particle acceleration, this also allows using larger electron beams (here: up to 800\,nm in diameter). Consequently, larger beam currents are permitted, as will be discussed in the next section.

\begin{table}
	\centering
	\renewcommand{\arraystretch}{1.2}
		 \begin{tabular}{|>{\centering}p{1cm}||>{\centering}p{1.5cm}|>{\centering}p{1.5cm}|>{\centering}p{1.5cm}|>{\centering}p{1.5cm}|}\hline
			\vspace{1ex}$\lambda$\,($\mu$m)											& \multicolumn{2}{>{\centering}p{3cm}|}{$E_\mathrm{kin}=29$\,keV\newline(Fig.\ \ref{fig:figure6}(d))} 	& \multicolumn{2}{>{\centering}p{3cm}|}{$E_\mathrm{kin}=957$\,keV\newline(Fig.\ \ref{fig:figure6}(g))}\tabularnewline
			\cline{2-5}
		 & 1\,GV/m & 10\,GV/m & 1\,GV/m & 10\,GV/m \tabularnewline
		\hline
		\hline
		0.8	& 	12\,$\mu$m				& 	4\,$\mu$m	&		149\,$\mu$m	&		47\,$\mu$m	 \tabularnewline
		\hline
		2 	& 		19\,$\mu$m			& 	6\,$\mu$m	&		236\,$\mu$m	&		75\,$\mu$m	 \tabularnewline
		\hline
		5 	& 		31\,$\mu$m			& 	10\,$\mu$m	&		373\,$\mu$m	&		118\,$\mu$m	 \tabularnewline
		\hline
		\end{tabular}
	\caption{Dephasing length $x_\mathrm{deph}$ for non-relativistic (29\,keV) and relativistic (957\,keV) electrons inside a double grating structure. The 29\,keV electrons pass the structure depicted in Fig.\ \ref{fig:figure6}\,(d) which is excited with a single laser with wavelength $\lambda$ and peak electric field $E_\mathrm{p}$. The 957\,keV electrons are accelerated inside the symmetrically pumped structure shown in Fig.\ \ref{fig:figure6}\,(g), excited by two lasers with $\lambda$ and $E_\mathrm{p}$. We assumed an acceleration gradient at the center of the vacuum channel of $G=0.027\cdot eE_\mathrm{p}$ for the non-relativistic case and $G=0.09\cdot eE_\mathrm{p}$ for the relativistic case. Hence, for an exciting wavelength of 2\,$\mu$m (5\,$\mu$m) the dimensions in Fig.\ \ref{fig:figure6} have to be scaled up by a factor of 2.5 (6.3), as discussed in the text.}
	\label{tab:dephasing}
\end{table}

In Table\ \ref{tab:dephasing} we list the dephasing lengths $x_\mathrm{deph}$ for the non-relativistic and relativistic electrons inside the double grating structures shown in Fig.\ \ref{fig:figure6}\,(d) and (g). We have calculated $x_\mathrm{deph}$ according to Eq.\ \ref{eq:deph11} using the maximum acceleration gradient $G_\mathrm{max}=ecC_\mathrm{s}/(\beta\gamma)$. For example, assuming $E_\mathrm{p}=1$\,GV/m, 30\,keV electrons that are accelerated inside the double grating structure of Fig.\ \ref{fig:figure6}\,(d) dephase after only 12\,$\mu$m. 1\,MeV electrons inside the structure, shown in Fig.\ \ref{fig:figure6}\,(g), stay in phase with the accelerating fields for about a factor of ten larger distances. Increasing the laser peak field reduces $x_\mathrm{deph}$ by a factor of $1/\sqrt{E_\mathrm{p}}$.

The dephasing length scales with the exciting laser wavelength $\lambda$ according to $\sqrt{\lambda}$. In the calculation of $x_\mathrm{deph}$ we assume the same acceleration efficiency for all wavelengths. This implies, first, that the refractive index of the grating material is constant, second, that the available field gradients (determined by material breakdown thresholds) are not significantly affected by the incident wavelength and, third, that the dimensions of the double grating structures have to be scaled up proportionally to the wavelength. Hence, for an exciting wavelength of 2\,$\mu$m the dimensions in Fig.\ \ref{fig:figure6} are increased by a factor of 2.5. For $\lambda=5$\,$\mu$m the dimensions are a factor of 6.3 larger.

%In Fig.\ \ref{fig:figure7} we show particle tracking results for the double grating structures with identical laser and electron parameters as used in Fig.\ \ref{fig:figure4}. It can be seen that the one-sided excitation for the 750\ nm double grating structure has a much lower efficiency than the single grating (Fig.\ \ref{fig:figure7}(d-f) vs. Fig.\ \ref{fig:figure4}(g-i)). Here, a two-sided excitation is preferable to maximize the acceleration gradient (Fig.\ \ref{fig:figure7}(g-i)).

\section{Space charge forces}

The transverse dimension of the particle beam inside an accelerator has to be smaller than the size of the accelerating structures, which is directly connected to the driving wavelength. While conventional RF accelerators can support beams with diameters on the cm-scale, optical linear accelerators need to be provided with particle beams with sub-micron diameters. Hence, space charge forces limit the maximum bunch charge of the particle ensemble especially in DLAs, because this repulsive force is inversely proportional to the transverse dimension of the beam and therefore more than four orders of magnitude larger in optical linacs as compared to RF accelerators. Elliptical or sheet beams with a large transverse dimension perpendicular to the vacuum channel have been suggested to minimize defocusing due to the space charge effect \cite{Rosenzweig1995}.

We use the paraxial ray equation to estimate the maximum bunch charge of a beam with a circular profile. We note that this approach is only a first step because it assumes a uniform beam density profile, which is only true for particle beams with a vanishing random transverse velocity spread \cite{Reiser2007}. The paraxial ray equation describes the beam envelope radius $r_{\mathrm{m}}$ via
\begin{equation}
\begin{split}
r_{\mathrm{m}}''&+\frac{\gamma'r_{\mathrm{m}}'}{\beta^2\gamma}+\frac{\gamma''r_{\mathrm{m}}}{2\beta^2\gamma}+\left(\frac{qB}{2mc\beta\gamma}\right)^2 r_{\mathrm{m}}\\
&-\left(\frac{p_{\theta}}{mc\beta\gamma}\right)^2\frac{1}{r_{\mathrm{m}}^3}-\frac{\epsilon_\mathrm{n}^2}{\beta^2\gamma^2r_{\mathrm{m}}^3}-\frac{K}{r_{\mathrm{m}}}=0.
\end{split}
\end{equation}
Here $B$ is an axial (static) magnetic field, $p_{\theta}$ the canonical angular momentum of the particles, $\epsilon_\mathrm{n}$ the normalized emittance and $K=2I/(I_0\beta^3\gamma^3)$ the generalized perveance, with the Budker or Alfv\'en current for electrons: $I_0=17000$\,A. The perveance is a measure for the space charge effect \cite{Lawson1958,Reiser2007}. The paraxial ray equation describes the beam dynamics and includes acceleration in a longitudinal electric field with $\gamma'=eE_{\|}/(mc^2)$ (2$^\mathrm{nd}$ term), focusing in a radial electric field with $\gamma''=2eE_\bot/(mc^2r_{\mathrm{m}})$ (3$^\mathrm{rd}$ term), focusing in an axial magnetic field (4$^\mathrm{th}$ term), defocusing due to angular momentum and normalized emittance (5$^\mathrm{th}$ and 6$^\mathrm{th}$ term), as well as defocusing due to space charge (last term).

To estimate the maximum bunch charge in an optical accelerator we assume $B=p_\theta=0$, an emittance-limited beam
($K\beta^2\gamma^2 r_{\mathrm{m}}^2 \ll \epsilon_n^{2}$) and focusing in the radial electric field provided by the optical accelerator. As can be seen from Eq.\ \ref{eq:int9} the focusing and accelerating fields are out of phase, that is, $\text{Re}\left(F_x\right)=0$ for $\left|\text{Re}\left(F_z\right)\right|=qc\left|B_y\right|/(\beta\gamma^2)$. Hence, for a focusing structure and a microbunched electron beam with sub-laser-cycle microbunch duration we can neglect acceleration, $\gamma'=0$. Therefore,
\begin{equation}
r_{\mathrm{m}}''+\frac{\gamma''r_{\mathrm{m}}}{2\beta^2\gamma}-\frac{\epsilon_\mathrm{n}^2}{\beta^2\gamma^2r_{\mathrm{m}}^3}=0.
\end{equation}
The acceleration gradient of the grating accelerator is defined as $G=dE_{\mathrm{kin}}/dx=\left|F_x\right|=qc\left|B_y\right|/(\beta\gamma)$. The transverse focusing force is equivalent to a radial electric field with amplitude $E_\bot=\left|F_z\right|/q=G/(q\gamma)$, thus
\begin{equation}
\gamma''=\frac{2qE_\bot}{mc^2r_{\mathrm{m}}}=\frac{2G}{mc^2r_{\mathrm{m}}\gamma}.
\end{equation}
Demanding a stable beam diameter, i.e.\ $r_{\mathrm{m}}''=0$, yields the emittance:
\begin{equation}
\epsilon_{\mathrm{n}}^2=\frac{2G r_{\mathrm{m}}^3}{mc^2}.
\end{equation}
To examine the typical emittance of this beam suited for DLAs, we take $G$ as 1 GeV/m and $r_{\mathrm{m}}=100$\,nm, yielding a normalized emittance of 2\,nm-rad. With $G$ as 10 GeV/m and $r_{\mathrm{m}}=300$\,nm, a normalized emittance of 32\,nm-rad is found. 
By assuming that the perveance term in the envelope equation is 10\% of the emittance term, effectively treating the perveance term as a perturbation, the maximum current allowed is determined:
\begin{equation}
I_{\mathrm{b}}=0.1I_0\frac{G\beta\gamma r_{\mathrm{m}}}{mc^2}.
\label{JoshCurrent}
\end{equation}
Note that the factor 0.1 in this equation derives from the assumption that the perveance term is 10\% of the emittance term. If one were to assume that the perveance term was only 1\% of the emittance term, the constant in Equation~\ref{JoshCurrent} would be 0.01 and all of the estimated currents in what follows would scale accordingly.
We calculate $I_\mathrm{b}$ for non-relativistic and relativistic electrons inside double grating structures based on our simulation results. We assume an electron beam with radius $r_\mathrm{m}$ propagating on-axis inside the structure. The acceleration gradient at the envelope of the beam is given by (Eq.\ \ref{eq:double1})
\begin{equation}
G=F_x|_{z=r_\mathrm{m}/2}=\frac{C_\mathrm{s}}{\beta\gamma}\cosh\left(\frac{k_0r_\mathrm{m}}{2\beta\gamma}\right).
\label{eq:spacecharge01}
\end{equation}
In Table\ \ref{tab:spacecharge} we show the maximum peak beam currents for a 30\,keV electron beam with $r_\mathrm{m}=50$\,nm, in the double grating structure (Figure \ref{fig:figure6}\,(d)), which is excited by a single laser, for various laser peak electric fields $E_\mathrm{p}$ and driving wavelengths $\lambda$. For example, for $E_\mathrm{p}=10$\,GV/m at $\lambda=800$\,nm this structure can sustain a peak beam current of $I_\mathrm{b}=18$\,mA. We further calculate $I_\mathrm{b}$ for a 1\,MeV electron beam, with $r_\mathrm{m}=300$\,nm, inside the symmetrically pumped structure depicted in Figure \ref{fig:figure6}\,(g). Here, the attainable peak currents are more than two orders of magnitude larger than in the non-relativistic case and a laser peak field $E_\mathrm{p}=7$\,GV/m implies $I_\mathrm{b}=2$\,A at 800\,nm and 12\,A at 5$\mu$m.

The maximum bunch charge that can be kept inside the accelerator without beam expansion due to space charge forces is
\begin{equation}
Q_{\mathrm{b}}=I_\mathrm{b}\tau_\mathrm{b},
\label{eq:spacecharge02}
\end{equation}
with the bunch duration $\tau_{\mathrm{b}}$. The bunch duration inside the accelerator is ideally smaller than the optical cycle $\tau_\mathrm{cycle}=\lambda/c$, so that the bunch experiences a homogeneous force. Assuming $\tau_\mathrm{b}=0.1\tau_\mathrm{cycle}$ and $\lambda=800$\,nm we obtain for the 30\,keV electrons and $E_\mathrm{p}=10$\,GV/m a maximum bunch charge of 4.8\,aC, corresponding to 30 electrons. For the 1\,MeV electrons and $E_\mathrm{p}=7$\,GV/m the structure can support 0.5\,fC at 800\,nm and 20 fC at 5\,$mu$m.

As the space charge force is inversely proportional to the beam radius, a laser wavelength of 2\,$\mu$m allows 2.5 times larger peak currents assuming that the grating dimensions and electron beam radius are scaled up in size accordingly. Moreover, we assume that the refractive index of the dielectric material is constant for all wavelengths, which assures the same excitation efficiency of the spatial harmonics. Similarly, a driving wavelength of 5\,$\mu$m allows a 6.3 times larger peak current. This implies that the bunch charge $Q_\mathrm{b}$ scales with $\lambda^2$, assuming that $\tau_\mathrm{b}$ is proportional to $\tau_\mathrm{cycle}$.

Note that in the derivation of the maximum current $I_\mathrm{b}$ we assume that the focusing force of the grating structure is used to counteract the defocusing force due to space charge, i.e., we choose a start phase of the microbunch for which the focusing force is maximum and acceleration is zero. For a useful accelerator structure transversal confinement in conjunction with longitudinal bunching and acceleration is required. As discussed above, this can be achieved, for example, with alternating phase focusing \cite{Good1953,Swenson1976,Wangler2008} or with biharmonic structures \cite{Naranjo2012}. Hence, the given numbers are estimates and might vary in a realistic setup. Furthermore, in the analysis above a circular transverse spot size was assumed; if instead the electron bunches were ribbon beams, larger currents could traverse the DLA. Regardless,  more involved simulations are needed to more realistically take into account space charge forces, wakefield effects, and other effects that are due to the nearby structure.

\begin{table}
	\centering
	\renewcommand{\arraystretch}{1.2}
		 \begin{tabular}{|>{\centering}p{1.3cm}||>{\centering}p{1cm}|>{\centering}p{1.05cm}|>{\centering}p{1.05cm}|>{\centering}p{1cm}|>{\centering}p{1cm}|>{\centering}p{1cm}|}\hline
			\multirow{5}{1.3cm}{\centering$E_\mathrm{p}$\,($\frac{\mathrm{GV}}{\mathrm{m}}$)}		 &\multicolumn{3}{>{\centering}p{3.1cm}|}{\centering$E_\mathrm{kin}=29$\,keV,\newline$r_\mathrm{m}=50$\,nm\newline(Fig.\ \ref{fig:figure6}\,(d))} 	& \multicolumn{3}{>{\centering}p{3cm}|}{$E_\mathrm{kin}=957$\,keV,\newline$r_\mathrm{m}=300$\,nm\newline(Fig.\ \ref{fig:figure6}\,(g))}\tabularnewline
		\cline{2-7}
		 & \multicolumn{3}{>{\centering}p{3.1cm}|}{$\lambda$\,($\mu$m)} 	& \multicolumn{3}{>{\centering}p{3cm}|}{$\lambda$\,($\mu$m)}\tabularnewline
		%\cline{2-7}
		& 0.8 & 2 & 5 & 0.8 & 2 & 5 \tabularnewline
		\hline
		\hline
		1	& 				1.8\,mA	& 4.4\,mA & 11.2\,mA						& 0.28\,A & 0.68\,A & 1.72\,A \tabularnewline
		\hline
		7 	& 				12.6\,mA & 32\,mA & 80\,mA						& 1.9\,A & 4.8\,A & 12\,A\tabularnewline
		\hline
		10 & 				18\,mA	& 46\,mA & 114\,mA					& 2.8\,A & 6.8\,A & 17.2\,A\tabularnewline
		\hline
		\end{tabular}
	\caption{Maximum peak beam current $I_\mathrm{b}$ for a 29\,keV electron beam ($r_\mathrm{m}=50$\,nm) inside the grating geometry shown in Fig.\ \ref{fig:figure6}\,(d), and for a 957\,keV electron beam ($r_\mathrm{m}=300$\,nm) inside the grating geometry depicted in Fig.\ \ref{fig:figure6}\,(g) for three different laser peak electric fields and the driving laser wavelengths 800\,nm, 2\,$\mu$m and 5\,$\mu$m.}
	\label{tab:spacecharge}
\end{table}

\section{conclusion and outlook}
In conclusion, we have derived the properties of the accelerating fields close to single and double dielectric grating structures. We have simulated the excitation of those fields and performed particle tracking simulations, which give a promising outlook for future dielectric laser accelerators. Our analysis of dephasing and space charge effects will be important for designing new structures. Next steps comprise the full simulation of electron bunches in double grating structures including wakefield losses, beam loading and space charge, determination of the optimal electron injection phase and the staging of multiple structures to correct for dephasing and deflection.

\section*{Acknowledgements}
We gratefully acknowledge H.\ Ramadas and R.\ Davies for simulation work, D.\ Dowell, K.\ Floettmann and the Stanford DARPA AXiS collaboration for discussions. This work has been funded by the DFG cluster of excellence Munich-Centre for Advanced Photonics.

\bibliography{library}

\begin{thebibliography}{42}%
\makeatletter
\providecommand \@ifxundefined [1]{%
 \@ifx{#1\undefined}
}%
\providecommand \@ifnum [1]{%
 \ifnum #1\expandafter \@firstoftwo
 \else \expandafter \@secondoftwo
 \fi
}%
\providecommand \@ifx [1]{%
 \ifx #1\expandafter \@firstoftwo
 \else \expandafter \@secondoftwo
 \fi
}%
\providecommand \natexlab [1]{#1}%
\providecommand \enquote  [1]{``#1''}%
\providecommand \bibnamefont  [1]{#1}%
\providecommand \bibfnamefont [1]{#1}%
\providecommand \citenamefont [1]{#1}%
\providecommand \href@noop [0]{\@secondoftwo}%
\providecommand \href [0]{\begingroup \@sanitize@url \@href}%
\providecommand \@href[1]{\@@startlink{#1}\@@href}%
\providecommand \@@href[1]{\endgroup#1\@@endlink}%
\providecommand \@sanitize@url [0]{\catcode `\\12\catcode `\$12\catcode
  `\&12\catcode `\#12\catcode `\^12\catcode `\_12\catcode `\%12\relax}%
\providecommand \@@startlink[1]{}%
\providecommand \@@endlink[0]{}%
\providecommand \url  [0]{\begingroup\@sanitize@url \@url }%
\providecommand \@url [1]{\endgroup\@href {#1}{\urlprefix }}%
\providecommand \urlprefix  [0]{URL }%
\providecommand \Eprint [0]{\href }%
\providecommand \doibase [0]{http://dx.doi.org/}%
\providecommand \selectlanguage [0]{\@gobble}%
\providecommand \bibinfo  [0]{\@secondoftwo}%
\providecommand \bibfield  [0]{\@secondoftwo}%
\providecommand \translation [1]{[#1]}%
\providecommand \BibitemOpen [0]{}%
\providecommand \bibitemStop [0]{}%
\providecommand \bibitemNoStop [0]{.\EOS\space}%
\providecommand \EOS [0]{\spacefactor3000\relax}%
\providecommand \BibitemShut  [1]{\csname bibitem#1\endcsname}%
\let\auto@bib@innerbib\@empty
%</preamble>
\bibitem [{\citenamefont {Shimoda}(1962)}]{Shimoda1962}%
  \BibitemOpen
  \bibfield  {author} {\bibinfo {author} {\bibfnamefont {K.}~\bibnamefont
  {Shimoda}},\ }\href {http://ao.osa.org/abstract.cfm?URI=ao-1-1-33} {\bibfield
   {journal} {\bibinfo  {journal} {Appl. Opt.}\ }\textbf {\bibinfo {volume}
  {1}},\ \bibinfo {pages} {33} (\bibinfo {year} {1962})}\BibitemShut {NoStop}%
\bibitem [{\citenamefont {Spataro}\ \emph {et~al.}(2011)\citenamefont
  {Spataro}, \citenamefont {Alesini}, \citenamefont {Chimenti}, \citenamefont
  {Dolgashev}, \citenamefont {Haase}, \citenamefont {Tantawi}, \citenamefont
  {Higashi}, \citenamefont {Marrelli}, \citenamefont {Mostacci}, \citenamefont
  {Parodi},\ and\ \citenamefont {Yeremian}}]{Spataro2011}%
  \BibitemOpen
  \bibfield  {author} {\bibinfo {author} {\bibfnamefont {B.}~\bibnamefont
  {Spataro}}, \bibinfo {author} {\bibfnamefont {D.}~\bibnamefont {Alesini}},
  \bibinfo {author} {\bibfnamefont {V.}~\bibnamefont {Chimenti}}, \bibinfo
  {author} {\bibfnamefont {V.}~\bibnamefont {Dolgashev}}, \bibinfo {author}
  {\bibfnamefont {A.}~\bibnamefont {Haase}}, \bibinfo {author} {\bibfnamefont
  {S.}~\bibnamefont {Tantawi}}, \bibinfo {author} {\bibfnamefont
  {Y.}~\bibnamefont {Higashi}}, \bibinfo {author} {\bibfnamefont
  {C.}~\bibnamefont {Marrelli}}, \bibinfo {author} {\bibfnamefont
  {A.}~\bibnamefont {Mostacci}}, \bibinfo {author} {\bibfnamefont
  {R.}~\bibnamefont {Parodi}}, \ and\ \bibinfo {author} {\bibfnamefont
  {A.}~\bibnamefont {Yeremian}},\ }\bibfield  {booktitle} {\emph {\bibinfo
  {booktitle} {X-Band Structures, Beam Dynamics and Sources Workshop
  (XB-10)}},\ }\href {\doibase 10.1016/j.nima.2011.05.020} {\bibfield
  {journal} {\bibinfo  {journal} {Nucl. Instrum. Methods A}\ }\textbf {\bibinfo
  {volume} {657}},\ \bibinfo {pages} {114} (\bibinfo {year}
  {2011})}\BibitemShut {NoStop}%
\bibitem [{\citenamefont {Solyak}(2009)}]{Solyak2009}%
  \BibitemOpen
  \bibfield  {author} {\bibinfo {author} {\bibfnamefont {N.~A.}\ \bibnamefont
  {Solyak}},\ }\href {http://dx.doi.org/10.1063/1.3080933} {\bibfield
  {journal} {\bibinfo  {journal} {AIP Conf. Proc.}\ }\textbf {\bibinfo {volume}
  {1086}},\ \bibinfo {pages} {365} (\bibinfo {year} {2009})}\BibitemShut
  {NoStop}%
\bibitem [{\citenamefont {Lenzner}\ \emph {et~al.}(1998)\citenamefont
  {Lenzner}, \citenamefont {Kr\"uger}, \citenamefont {Sartania}, \citenamefont
  {Cheng}, \citenamefont {Spielmann}, \citenamefont {Mourou}, \citenamefont
  {Kautek},\ and\ \citenamefont {Krausz}}]{lenzner1998}%
  \BibitemOpen
  \bibfield  {author} {\bibinfo {author} {\bibfnamefont {M.}~\bibnamefont
  {Lenzner}}, \bibinfo {author} {\bibfnamefont {J.}~\bibnamefont {Kr\"uger}},
  \bibinfo {author} {\bibfnamefont {S.}~\bibnamefont {Sartania}}, \bibinfo
  {author} {\bibfnamefont {Z.}~\bibnamefont {Cheng}}, \bibinfo {author}
  {\bibfnamefont {C.}~\bibnamefont {Spielmann}}, \bibinfo {author}
  {\bibfnamefont {G.}~\bibnamefont {Mourou}}, \bibinfo {author} {\bibfnamefont
  {W.}~\bibnamefont {Kautek}}, \ and\ \bibinfo {author} {\bibfnamefont
  {F.}~\bibnamefont {Krausz}},\ }\href {\doibase 10.1103/PhysRevLett.80.4076}
  {\bibfield  {journal} {\bibinfo  {journal} {Phys. Rev. Lett.}\ }\textbf
  {\bibinfo {volume} {80}},\ \bibinfo {pages} {4076} (\bibinfo {year}
  {1998})}\BibitemShut {NoStop}%
\bibitem [{\citenamefont {Rosenzweig}\ \emph {et~al.}(1995)\citenamefont
  {Rosenzweig}, \citenamefont {Murokh},\ and\ \citenamefont
  {Pellegrini}}]{Rosenzweig1995}%
  \BibitemOpen
  \bibfield  {author} {\bibinfo {author} {\bibfnamefont {J.}~\bibnamefont
  {Rosenzweig}}, \bibinfo {author} {\bibfnamefont {A.}~\bibnamefont {Murokh}},
  \ and\ \bibinfo {author} {\bibfnamefont {C.}~\bibnamefont {Pellegrini}},\
  }\href {http://link.aps.org/doi/10.1103/PhysRevLett.74.2467} {\bibfield
  {journal} {\bibinfo  {journal} {Phys. Rev. Lett.}\ }\textbf {\bibinfo
  {volume} {74}},\ \bibinfo {pages} {2467} (\bibinfo {year}
  {1995})}\BibitemShut {NoStop}%
\bibitem [{\citenamefont {Huang}\ \emph {et~al.}(1996)\citenamefont {Huang},
  \citenamefont {Zheng}, \citenamefont {Tulloch},\ and\ \citenamefont
  {Byer}}]{Huang1996}%
  \BibitemOpen
  \bibfield  {author} {\bibinfo {author} {\bibfnamefont {Y.~C.}\ \bibnamefont
  {Huang}}, \bibinfo {author} {\bibfnamefont {D.}~\bibnamefont {Zheng}},
  \bibinfo {author} {\bibfnamefont {W.~M.}\ \bibnamefont {Tulloch}}, \ and\
  \bibinfo {author} {\bibfnamefont {R.~L.}\ \bibnamefont {Byer}},\ }\href
  {http://dx.doi.org/10.1063/1.116731} {\bibfield  {journal} {\bibinfo
  {journal} {Appl. Phys. Lett.}\ }\textbf {\bibinfo {volume} {68}},\ \bibinfo
  {pages} {753} (\bibinfo {year} {1996})}\BibitemShut {NoStop}%
\bibitem [{\citenamefont {Lawson}(1979)}]{Lawson1979}%
  \BibitemOpen
  \bibfield  {author} {\bibinfo {author} {\bibfnamefont {J.~D.}\ \bibnamefont
  {Lawson}},\ }\href {\doibase http://dx.doi.org/10.1109/TNS.1979.4330749}
  {\bibfield  {journal} {\bibinfo  {journal} {IEEE Trans. Nucl. Sci.}\ }\textbf
  {\bibinfo {volume} {26}},\ \bibinfo {pages} {4217} (\bibinfo {year}
  {1979})}\BibitemShut {NoStop}%
\bibitem [{\citenamefont {Scully}\ and\ \citenamefont
  {Zubairy}(1991)}]{Scully1991}%
  \BibitemOpen
  \bibfield  {author} {\bibinfo {author} {\bibfnamefont {M.~O.}\ \bibnamefont
  {Scully}}\ and\ \bibinfo {author} {\bibfnamefont {M.~S.}\ \bibnamefont
  {Zubairy}},\ }\href {http://link.aps.org/doi/10.1103/PhysRevA.44.2656}
  {\bibfield  {journal} {\bibinfo  {journal} {Phys. Rev. A}\ }\textbf {\bibinfo
  {volume} {44}},\ \bibinfo {pages} {2656} (\bibinfo {year}
  {1991})}\BibitemShut {NoStop}%
\bibitem [{\citenamefont {Esarey}\ \emph {et~al.}(2009)\citenamefont {Esarey},
  \citenamefont {Schroeder},\ and\ \citenamefont {Leemans}}]{esarey2009}%
  \BibitemOpen
  \bibfield  {author} {\bibinfo {author} {\bibfnamefont {E.}~\bibnamefont
  {Esarey}}, \bibinfo {author} {\bibfnamefont {C.~B.}\ \bibnamefont
  {Schroeder}}, \ and\ \bibinfo {author} {\bibfnamefont {W.~P.}\ \bibnamefont
  {Leemans}},\ }\href {\doibase 10.1103/RevModPhys.81.1229} {\bibfield
  {journal} {\bibinfo  {journal} {Rev. Mod. Phys.}\ }\textbf {\bibinfo {volume}
  {81}},\ \bibinfo {pages} {1229} (\bibinfo {year} {2009})}\BibitemShut
  {NoStop}%
\bibitem [{\citenamefont {Takeda}\ and\ \citenamefont
  {Matsui}(1968)}]{Takeda1968}%
  \BibitemOpen
  \bibfield  {author} {\bibinfo {author} {\bibfnamefont {Y.}~\bibnamefont
  {Takeda}}\ and\ \bibinfo {author} {\bibfnamefont {I.}~\bibnamefont
  {Matsui}},\ }\href {\doibase 10.1016/0029-554X(68)90378-9} {\bibfield
  {journal} {\bibinfo  {journal} {Nucl. Instrum. Methods}\ }\textbf {\bibinfo
  {volume} {62}},\ \bibinfo {pages} {306} (\bibinfo {year} {1968})}\BibitemShut
  {NoStop}%
\bibitem [{\citenamefont {Palmer}(1980)}]{palmer1980}%
  \BibitemOpen
  \bibfield  {author} {\bibinfo {author} {\bibfnamefont {R.}~\bibnamefont
  {Palmer}},\ }\href@noop {} {\bibfield  {journal} {\bibinfo  {journal} {Part.
  Accel.}\ }\textbf {\bibinfo {volume} {11}},\ \bibinfo {pages} {81} (\bibinfo
  {year} {1980})}\BibitemShut {NoStop}%
\bibitem [{\citenamefont {Mizuno}\ \emph {et~al.}(1975)\citenamefont {Mizuno},
  \citenamefont {Ono},\ and\ \citenamefont {Shimoe}}]{Mizuno1975}%
  \BibitemOpen
  \bibfield  {author} {\bibinfo {author} {\bibfnamefont {K.}~\bibnamefont
  {Mizuno}}, \bibinfo {author} {\bibfnamefont {S.}~\bibnamefont {Ono}}, \ and\
  \bibinfo {author} {\bibfnamefont {O.}~\bibnamefont {Shimoe}},\ }\href
  {http://dx.doi.org/10.1038/253184a0} {\bibfield  {journal} {\bibinfo
  {journal} {Nature}\ }\textbf {\bibinfo {volume} {253}},\ \bibinfo {pages}
  {184} (\bibinfo {year} {1975})}\BibitemShut {NoStop}%
\bibitem [{\citenamefont {Mizuno}\ \emph {et~al.}(1987)\citenamefont {Mizuno},
  \citenamefont {Pae}, \citenamefont {Nozokido},\ and\ \citenamefont
  {Furuya}}]{Mizuno1987}%
  \BibitemOpen
  \bibfield  {author} {\bibinfo {author} {\bibfnamefont {K.}~\bibnamefont
  {Mizuno}}, \bibinfo {author} {\bibfnamefont {J.}~\bibnamefont {Pae}},
  \bibinfo {author} {\bibfnamefont {T.}~\bibnamefont {Nozokido}}, \ and\
  \bibinfo {author} {\bibfnamefont {K.}~\bibnamefont {Furuya}},\ }\href
  {http://dx.doi.org/10.1038/328045a0} {\bibfield  {journal} {\bibinfo
  {journal} {Nature}\ }\textbf {\bibinfo {volume} {328}},\ \bibinfo {pages}
  {45} (\bibinfo {year} {1987})}\BibitemShut {NoStop}%
\bibitem [{\citenamefont {Bae}\ \emph {et~al.}(1992)\citenamefont {Bae},
  \citenamefont {Shirai}, \citenamefont {Nishida}, \citenamefont {Nozokido},
  \citenamefont {Furuya},\ and\ \citenamefont {Mizuno}}]{Bae1992}%
  \BibitemOpen
  \bibfield  {author} {\bibinfo {author} {\bibfnamefont {J.}~\bibnamefont
  {Bae}}, \bibinfo {author} {\bibfnamefont {H.}~\bibnamefont {Shirai}},
  \bibinfo {author} {\bibfnamefont {T.}~\bibnamefont {Nishida}}, \bibinfo
  {author} {\bibfnamefont {T.}~\bibnamefont {Nozokido}}, \bibinfo {author}
  {\bibfnamefont {K.}~\bibnamefont {Furuya}}, \ and\ \bibinfo {author}
  {\bibfnamefont {K.}~\bibnamefont {Mizuno}},\ }\href
  {http://dx.doi.org/10.1063/1.107773} {\bibfield  {journal} {\bibinfo
  {journal} {Appl. Phys. Lett.}\ }\textbf {\bibinfo {volume} {61}},\ \bibinfo
  {pages} {870} (\bibinfo {year} {1992})}\BibitemShut {NoStop}%
\bibitem [{\citenamefont {Plettner}\ \emph
  {et~al.}(2005{\natexlab{a}})\citenamefont {Plettner}, \citenamefont {Byer},
  \citenamefont {Colby}, \citenamefont {Cowan}, \citenamefont {Sears},
  \citenamefont {Spencer},\ and\ \citenamefont {Siemann}}]{plettner2005b}%
  \BibitemOpen
  \bibfield  {author} {\bibinfo {author} {\bibfnamefont {T.}~\bibnamefont
  {Plettner}}, \bibinfo {author} {\bibfnamefont {R.~L.}\ \bibnamefont {Byer}},
  \bibinfo {author} {\bibfnamefont {E.}~\bibnamefont {Colby}}, \bibinfo
  {author} {\bibfnamefont {B.}~\bibnamefont {Cowan}}, \bibinfo {author}
  {\bibfnamefont {C.~M.~S.}\ \bibnamefont {Sears}}, \bibinfo {author}
  {\bibfnamefont {J.~E.}\ \bibnamefont {Spencer}}, \ and\ \bibinfo {author}
  {\bibfnamefont {R.~H.}\ \bibnamefont {Siemann}},\ }\href
  {http://link.aps.org/doi/10.1103/PhysRevLett.95.134801} {\bibfield  {journal}
  {\bibinfo  {journal} {Phys. Rev. Lett.}\ }\textbf {\bibinfo {volume} {95}},\
  \bibinfo {pages} {134801} (\bibinfo {year} {2005}{\natexlab{a}})}\BibitemShut
  {NoStop}%
\bibitem [{\citenamefont {Plettner}\ \emph
  {et~al.}(2005{\natexlab{b}})\citenamefont {Plettner}, \citenamefont {Byer},
  \citenamefont {Colby}, \citenamefont {Cowan}, \citenamefont {Sears},
  \citenamefont {Spencer},\ and\ \citenamefont {Siemann}}]{plettner2005}%
  \BibitemOpen
  \bibfield  {author} {\bibinfo {author} {\bibfnamefont {T.}~\bibnamefont
  {Plettner}}, \bibinfo {author} {\bibfnamefont {R.~L.}\ \bibnamefont {Byer}},
  \bibinfo {author} {\bibfnamefont {E.}~\bibnamefont {Colby}}, \bibinfo
  {author} {\bibfnamefont {B.}~\bibnamefont {Cowan}}, \bibinfo {author}
  {\bibfnamefont {C.~M.~S.}\ \bibnamefont {Sears}}, \bibinfo {author}
  {\bibfnamefont {J.~E.}\ \bibnamefont {Spencer}}, \ and\ \bibinfo {author}
  {\bibfnamefont {R.~H.}\ \bibnamefont {Siemann}},\ }\href {\doibase
  10.1103/PhysRevSTAB.8.121301} {\bibfield  {journal} {\bibinfo  {journal}
  {Phys. Rev. ST Accel. Beams}\ }\textbf {\bibinfo {volume} {8}},\ \bibinfo
  {pages} {121301} (\bibinfo {year} {2005}{\natexlab{b}})}\BibitemShut
  {NoStop}%
\bibitem [{\citenamefont {Plettner}\ \emph {et~al.}(2006)\citenamefont
  {Plettner}, \citenamefont {Lu},\ and\ \citenamefont {Byer}}]{plettner2006}%
  \BibitemOpen
  \bibfield  {author} {\bibinfo {author} {\bibfnamefont {T.}~\bibnamefont
  {Plettner}}, \bibinfo {author} {\bibfnamefont {P.~P.}\ \bibnamefont {Lu}}, \
  and\ \bibinfo {author} {\bibfnamefont {R.~L.}\ \bibnamefont {Byer}},\ }\href
  {\doibase 10.1103/PhysRevSTAB.9.111301} {\bibfield  {journal} {\bibinfo
  {journal} {Phys. Rev. ST Accel. Beams}\ }\textbf {\bibinfo {volume} {9}},\
  \bibinfo {pages} {111301} (\bibinfo {year} {2006})}\BibitemShut {NoStop}%
\bibitem [{\citenamefont {Plettner}\ and\ \citenamefont
  {Byer}(2008{\natexlab{a}})}]{plettner2008}%
  \BibitemOpen
  \bibfield  {author} {\bibinfo {author} {\bibfnamefont {T.}~\bibnamefont
  {Plettner}}\ and\ \bibinfo {author} {\bibfnamefont {R.~L.}\ \bibnamefont
  {Byer}},\ }\href {\doibase 10.1103/PhysRevSTAB.11.030704} {\bibfield
  {journal} {\bibinfo  {journal} {Phys. Rev. ST Accel. Beams}\ }\textbf
  {\bibinfo {volume} {11}},\ \bibinfo {pages} {030704} (\bibinfo {year}
  {2008}{\natexlab{a}})}\BibitemShut {NoStop}%
\bibitem [{\citenamefont {Plettner}\ and\ \citenamefont
  {Byer}(2008{\natexlab{b}})}]{plettner2008b}%
  \BibitemOpen
  \bibfield  {author} {\bibinfo {author} {\bibfnamefont {T.}~\bibnamefont
  {Plettner}}\ and\ \bibinfo {author} {\bibfnamefont {R.~L.}\ \bibnamefont
  {Byer}},\ }\href {\doibase DOI: 10.1016/j.nima.2008.04.063} {\bibfield
  {journal} {\bibinfo  {journal} {Nucl. Instrum. Methods A}\ }\textbf {\bibinfo
  {volume} {593}},\ \bibinfo {pages} {63} (\bibinfo {year}
  {2008}{\natexlab{b}})}\BibitemShut {NoStop}%
\bibitem [{\citenamefont {Plettner}\ \emph {et~al.}(2009)\citenamefont
  {Plettner}, \citenamefont {Byer}, \citenamefont {McGuinness},\ and\
  \citenamefont {Hommelhoff}}]{plettner2009}%
  \BibitemOpen
  \bibfield  {author} {\bibinfo {author} {\bibfnamefont {T.}~\bibnamefont
  {Plettner}}, \bibinfo {author} {\bibfnamefont {R.~L.}\ \bibnamefont {Byer}},
  \bibinfo {author} {\bibfnamefont {C.}~\bibnamefont {McGuinness}}, \ and\
  \bibinfo {author} {\bibfnamefont {P.}~\bibnamefont {Hommelhoff}},\ }\href
  {\doibase 10.1103/PhysRevSTAB.12.101302} {\bibfield  {journal} {\bibinfo
  {journal} {Phys. Rev. ST Accel. Beams}\ }\textbf {\bibinfo {volume} {12}},\
  \bibinfo {pages} {101302} (\bibinfo {year} {2009})}\BibitemShut {NoStop}%
\bibitem [{\citenamefont {Plettner}\ \emph {et~al.}(2011)\citenamefont
  {Plettner}, \citenamefont {Byer},\ and\ \citenamefont
  {Montazeri}}]{Plettner2011}%
  \BibitemOpen
  \bibfield  {author} {\bibinfo {author} {\bibfnamefont {T.}~\bibnamefont
  {Plettner}}, \bibinfo {author} {\bibfnamefont {R.~L.}\ \bibnamefont {Byer}},
  \ and\ \bibinfo {author} {\bibfnamefont {B.}~\bibnamefont {Montazeri}},\
  }\href {http://dx.doi.org/10.1080/09500340.2011.611914} {\bibfield  {journal}
  {\bibinfo  {journal} {J. Mod. Opt.}\ }\textbf {\bibinfo {volume} {58}},\
  \bibinfo {pages} {1518} (\bibinfo {year} {2011})}\BibitemShut {NoStop}%
\bibitem [{\citenamefont {Yoder}\ and\ \citenamefont
  {Rosenzweig}(2005)}]{Yoder2005}%
  \BibitemOpen
  \bibfield  {author} {\bibinfo {author} {\bibfnamefont {R.~B.}\ \bibnamefont
  {Yoder}}\ and\ \bibinfo {author} {\bibfnamefont {J.~B.}\ \bibnamefont
  {Rosenzweig}},\ }\href {http://link.aps.org/doi/10.1103/PhysRevSTAB.8.111301}
  {\bibfield  {journal} {\bibinfo  {journal} {Phys. Rev. ST Accel. Beams}\
  }\textbf {\bibinfo {volume} {8}},\ \bibinfo {pages} {111301} (\bibinfo {year}
  {2005})}\BibitemShut {NoStop}%
\bibitem [{\citenamefont {Cowan}(2008)}]{Cowan2008}%
  \BibitemOpen
  \bibfield  {author} {\bibinfo {author} {\bibfnamefont {B.~M.}\ \bibnamefont
  {Cowan}},\ }\href {http://link.aps.org/doi/10.1103/PhysRevSTAB.11.011301}
  {\bibfield  {journal} {\bibinfo  {journal} {Phys. Rev. ST Accel. Beams}\
  }\textbf {\bibinfo {volume} {11}},\ \bibinfo {pages} {011301} (\bibinfo
  {year} {2008})}\BibitemShut {NoStop}%
\bibitem [{\citenamefont {Naranjo}\ \emph {et~al.}(2012)\citenamefont
  {Naranjo}, \citenamefont {Valloni}, \citenamefont {Putterman},\ and\
  \citenamefont {Rosenzweig}}]{Naranjo2012}%
  \BibitemOpen
  \bibfield  {author} {\bibinfo {author} {\bibfnamefont {B.}~\bibnamefont
  {Naranjo}}, \bibinfo {author} {\bibfnamefont {A.}~\bibnamefont {Valloni}},
  \bibinfo {author} {\bibfnamefont {S.}~\bibnamefont {Putterman}}, \ and\
  \bibinfo {author} {\bibfnamefont {J.~B.}\ \bibnamefont {Rosenzweig}},\ }\href
  {http://link.aps.org/doi/10.1103/PhysRevLett.109.164803} {\bibfield
  {journal} {\bibinfo  {journal} {Phys. Rev. Lett.}\ }\textbf {\bibinfo
  {volume} {109}},\ \bibinfo {pages} {164803} (\bibinfo {year}
  {2012})}\BibitemShut {NoStop}%
\bibitem [{\citenamefont {Breuer}\ and\ \citenamefont
  {Hommelhoff}(2013)}]{Breuer2013}%
  \BibitemOpen
  \bibfield  {author} {\bibinfo {author} {\bibfnamefont {J.}~\bibnamefont
  {Breuer}}\ and\ \bibinfo {author} {\bibfnamefont {P.}~\bibnamefont
  {Hommelhoff}},\ }\href
  {http://link.aps.org/doi/10.1103/PhysRevLett.111.134803} {\bibfield
  {journal} {\bibinfo  {journal} {Phys. Rev. Lett.}\ }\textbf {\bibinfo
  {volume} {111}},\ \bibinfo {pages} {134803} (\bibinfo {year}
  {2013})}\BibitemShut {NoStop}%
\bibitem [{\citenamefont {Peralta}\ \emph {et~al.}(2664)\citenamefont
  {Peralta}, \citenamefont {Soong}, \citenamefont {England}, \citenamefont
  {Colby}, \citenamefont {Wu}, \citenamefont {Montazeri}, \citenamefont
  {McGuinness}, \citenamefont {McNeur}, \citenamefont {Leedle}, \citenamefont
  {Walz}, \citenamefont {Sozer}, \citenamefont {Cowan}, \citenamefont
  {Schwartz}, \citenamefont {Travish},\ and\ \citenamefont
  {Byer}}]{Peralta2013}%
  \BibitemOpen
  \bibfield  {author} {\bibinfo {author} {\bibfnamefont {E.~A.}\ \bibnamefont
  {Peralta}}, \bibinfo {author} {\bibfnamefont {K.}~\bibnamefont {Soong}},
  \bibinfo {author} {\bibfnamefont {R.~J.}\ \bibnamefont {England}}, \bibinfo
  {author} {\bibfnamefont {E.~R.}\ \bibnamefont {Colby}}, \bibinfo {author}
  {\bibfnamefont {Z.}~\bibnamefont {Wu}}, \bibinfo {author} {\bibfnamefont
  {B.}~\bibnamefont {Montazeri}}, \bibinfo {author} {\bibfnamefont
  {C.}~\bibnamefont {McGuinness}}, \bibinfo {author} {\bibfnamefont
  {J.}~\bibnamefont {McNeur}}, \bibinfo {author} {\bibfnamefont {K.~J.}\
  \bibnamefont {Leedle}}, \bibinfo {author} {\bibfnamefont {D.}~\bibnamefont
  {Walz}}, \bibinfo {author} {\bibfnamefont {E.~B.}\ \bibnamefont {Sozer}},
  \bibinfo {author} {\bibfnamefont {B.}~\bibnamefont {Cowan}}, \bibinfo
  {author} {\bibfnamefont {B.}~\bibnamefont {Schwartz}}, \bibinfo {author}
  {\bibfnamefont {G.}~\bibnamefont {Travish}}, \ and\ \bibinfo {author}
  {\bibfnamefont {R.~L.}\ \bibnamefont {Byer}},\ }\href
  {http://dx.doi.org/10.1038/nature12664} {\bibfield  {journal} {\bibinfo
  {journal} {Nature}\ }\textbf {\bibinfo {volume} {advance online
  publication}},\ \bibinfo {pages} {27 Sept 2013} (\bibinfo {year} {DOI
  10.1038/nature12664})}\BibitemShut {NoStop}%
\bibitem [{\citenamefont {Palmer}(1988)}]{PALMER1988}%
  \BibitemOpen
  \bibfield  {author} {\bibinfo {author} {\bibfnamefont {R.}~\bibnamefont
  {Palmer}},\ }\href {http://dx.doi.org/10.1007/BFb0031508} {\bibfield
  {journal} {\bibinfo  {journal} {Frontiers of Particle Beams}\ }\textbf
  {\bibinfo {volume} {296}},\ \bibinfo {pages} {607} (\bibinfo {year}
  {1988})}\BibitemShut {NoStop}%
\bibitem [{\citenamefont {Good}(1953)}]{Good1953}%
  \BibitemOpen
  \bibfield  {author} {\bibinfo {author} {\bibfnamefont {M.~L.}\ \bibnamefont
  {Good}},\ }\href@noop {} {\bibfield  {journal} {\bibinfo  {journal} {Phys.
  Rev.}\ }\textbf {\bibinfo {volume} {92}},\ \bibinfo {pages} {538} (\bibinfo
  {year} {1953})}\BibitemShut {NoStop}%
\bibitem [{\citenamefont {Swenson}(1976)}]{Swenson1976}%
  \BibitemOpen
  \bibfield  {author} {\bibinfo {author} {\bibfnamefont {D.~A.}\ \bibnamefont
  {Swenson}},\ }\href {http://cdsweb.cern.ch/record/1053127/files/p61.pdf}
  {\bibfield  {journal} {\bibinfo  {journal} {Part. Accel.}\ }\textbf {\bibinfo
  {volume} {7}},\ \bibinfo {pages} {61} (\bibinfo {year} {1976})}\BibitemShut
  {NoStop}%
\bibitem [{\citenamefont {Wangler}(2008)}]{Wangler2008}%
  \BibitemOpen
  \bibfield  {author} {\bibinfo {author} {\bibfnamefont {T.~P.}\ \bibnamefont
  {Wangler}},\ }\href@noop {} {\emph {\bibinfo {title} {RF Linear
  accelerators}}}\ (\bibinfo  {publisher} {Wiley-VCH},\ \bibinfo {year}
  {2008})\BibitemShut {NoStop}%
\bibitem [{\citenamefont {Yee}(1966)}]{yee1966}%
  \BibitemOpen
  \bibfield  {author} {\bibinfo {author} {\bibfnamefont {K.}~\bibnamefont
  {Yee}},\ }\href {\doibase 10.1109/TAP.1966.1138693} {\bibfield  {journal}
  {\bibinfo  {journal} {IEEE Trans. Antennas Propag.}\ }\textbf {\bibinfo
  {volume} {14}},\ \bibinfo {pages} {302} (\bibinfo {year} {1966})}\BibitemShut
  {NoStop}%
\bibitem [{\citenamefont {Coccioli}\ \emph {et~al.}(1996)\citenamefont
  {Coccioli}, \citenamefont {Itoh}, \citenamefont {Pelosi},\ and\ \citenamefont
  {Silvester}}]{Coccioli1996}%
  \BibitemOpen
  \bibfield  {author} {\bibinfo {author} {\bibfnamefont {R.}~\bibnamefont
  {Coccioli}}, \bibinfo {author} {\bibfnamefont {T.}~\bibnamefont {Itoh}},
  \bibinfo {author} {\bibfnamefont {G.}~\bibnamefont {Pelosi}}, \ and\ \bibinfo
  {author} {\bibfnamefont {P.}~\bibnamefont {Silvester}},\ }\href {\doibase
  http://dx.doi.org/10.1109/74.556518} {\bibfield  {journal} {\bibinfo
  {journal} {IEEE Trans. Antennas Propag.}\ }\textbf {\bibinfo {volume} {38}},\
  \bibinfo {pages} {34} (\bibinfo {year} {1996})}\BibitemShut {NoStop}%
\bibitem [{\citenamefont {Ferrari}(2007)}]{Ferrari2007}%
  \BibitemOpen
  \bibfield  {author} {\bibinfo {author} {\bibfnamefont {R.}~\bibnamefont
  {Ferrari}},\ }\href {\doibase 10.1109/MAP.2007.4293978} {\bibfield  {journal}
  {\bibinfo  {journal} {IEEE Trans. Antennas Propag.}\ }\textbf {\bibinfo
  {volume} {49}},\ \bibinfo {pages} {216} (\bibinfo {year} {2007})}\BibitemShut
  {NoStop}%
\bibitem [{\citenamefont {Weiland}(1977)}]{Weiland1977}%
  \BibitemOpen
  \bibfield  {author} {\bibinfo {author} {\bibfnamefont {T.}~\bibnamefont
  {Weiland}},\ }\href@noop {} {\bibfield  {journal} {\bibinfo  {journal}
  {Archiv Elektronik und \"Ubertragungstechnik}\ }\textbf {\bibinfo {volume}
  {31}},\ \bibinfo {pages} {116} (\bibinfo {year} {1977})}\BibitemShut
  {NoStop}%
\bibitem [{\citenamefont {Liu}\ and\ \citenamefont {Zhao}(2005)}]{Liu2005}%
  \BibitemOpen
  \bibfield  {author} {\bibinfo {author} {\bibfnamefont {Q.~H.}\ \bibnamefont
  {Liu}}\ and\ \bibinfo {author} {\bibfnamefont {G.}~\bibnamefont {Zhao}},\
  }in\ \href@noop {} {\emph {\bibinfo {booktitle} {Computational
  Electromagnetics: The Finite-Difference Time-Domain Method}}},\ \bibinfo
  {editor} {edited by\ \bibinfo {editor} {\bibfnamefont {A.}~\bibnamefont
  {Taflove}}\ and\ \bibinfo {editor} {\bibfnamefont {S.}~\bibnamefont
  {Hagness}}}\ (\bibinfo  {publisher} {Artech House, Inc.},\ \bibinfo {year}
  {2005})\BibitemShut {NoStop}%
\bibitem [{\citenamefont {Pai}\ and\ \citenamefont {Awada}(1991)}]{Pai1991}%
  \BibitemOpen
  \bibfield  {author} {\bibinfo {author} {\bibfnamefont {D.~M.}\ \bibnamefont
  {Pai}}\ and\ \bibinfo {author} {\bibfnamefont {K.~A.}\ \bibnamefont
  {Awada}},\ }\href {http://josaa.osa.org/abstract.cfm?URI=josaa-8-5-755}
  {\bibfield  {journal} {\bibinfo  {journal} {J. Opt. Soc. Am. A}\ }\textbf
  {\bibinfo {volume} {8}},\ \bibinfo {pages} {755} (\bibinfo {year}
  {1991})}\BibitemShut {NoStop}%
\bibitem [{\citenamefont {Tremain}\ and\ \citenamefont
  {Mei}(1978)}]{Tremain1978}%
  \BibitemOpen
  \bibfield  {author} {\bibinfo {author} {\bibfnamefont {D.~E.}\ \bibnamefont
  {Tremain}}\ and\ \bibinfo {author} {\bibfnamefont {K.~K.}\ \bibnamefont
  {Mei}},\ }\href
  {http://www.opticsinfobase.org/abstract.cfm?URI=josa-68-6-775} {\bibfield
  {journal} {\bibinfo  {journal} {J. Opt. Soc. Am.}\ }\textbf {\bibinfo
  {volume} {68}},\ \bibinfo {pages} {775} (\bibinfo {year} {1978})}\BibitemShut
  {NoStop}%
\bibitem [{\citenamefont {Soong}\ \emph {et~al.}(2011)\citenamefont {Soong},
  \citenamefont {Byer}, \citenamefont {McGuinness}, \citenamefont {Peralta},\
  and\ \citenamefont {Colby}}]{Soong2011b}%
  \BibitemOpen
  \bibfield  {author} {\bibinfo {author} {\bibfnamefont {K.}~\bibnamefont
  {Soong}}, \bibinfo {author} {\bibfnamefont {R.~L.}\ \bibnamefont {Byer}},
  \bibinfo {author} {\bibfnamefont {C.}~\bibnamefont {McGuinness}}, \bibinfo
  {author} {\bibfnamefont {E.}~\bibnamefont {Peralta}}, \ and\ \bibinfo
  {author} {\bibfnamefont {E.}~\bibnamefont {Colby}},\ }in\ \href@noop {}
  {\emph {\bibinfo {booktitle} {Proceedings of 2011 Particle Accelerator
  Conference}}}\ (\bibinfo  {publisher} {IEEE},\ \bibinfo {address} {New
  York},\ \bibinfo {year} {2011})\ pp.\ \bibinfo {pages} {277--279}\BibitemShut
  {NoStop}%
\bibitem [{Note1()}]{Note1}%
  \BibitemOpen
  \bibinfo {note} {Note that the speed-of-light mode, which only occurs for
  $\lambda _\protect \mathrm {p}=\lambda $, is not included in this simulation.
  However, it can be included as shown in \cite {Plettner2011}. Further
  simulations are needed to investigate how a potentially excited
  speed-of-light mode affects the acceleration gradient of 1\protect \tmspace
  +\thinmuskip {.1667em}MeV electrons.}\BibitemShut {Stop}%
\bibitem [{\citenamefont {Aimidula}\ \emph {et~al.}(2014)\citenamefont
  {Aimidula}, \citenamefont {Welsch}, \citenamefont {Bake}, \citenamefont
  {Wan}, \citenamefont {Xie}, \citenamefont {Mete}, \citenamefont {Ueseka},
  \citenamefont {Matsumara}, \citenamefont {Yoshida},\ and\ \citenamefont
  {Koyama}}]{Aimidula2014}%
  \BibitemOpen
  \bibfield  {author} {\bibinfo {author} {\bibfnamefont {A.}~\bibnamefont
  {Aimidula}}, \bibinfo {author} {\bibfnamefont {C.}~\bibnamefont {Welsch}},
  \bibinfo {author} {\bibfnamefont {M.}~\bibnamefont {Bake}}, \bibinfo {author}
  {\bibfnamefont {F.}~\bibnamefont {Wan}}, \bibinfo {author} {\bibfnamefont
  {B.}~\bibnamefont {Xie}}, \bibinfo {author} {\bibfnamefont {O.}~\bibnamefont
  {Mete}}, \bibinfo {author} {\bibfnamefont {M.}~\bibnamefont {Ueseka}},
  \bibinfo {author} {\bibfnamefont {Y.}~\bibnamefont {Matsumara}}, \bibinfo
  {author} {\bibfnamefont {M.}~\bibnamefont {Yoshida}}, \ and\ \bibinfo
  {author} {\bibfnamefont {K.}~\bibnamefont {Koyama}},\ }\href@noop {}
  {\bibfield  {journal} {\bibinfo  {journal} {Phys. Plasmas}\ }\textbf
  {\bibinfo {volume} {21}},\ \bibinfo {pages} {023110} (\bibinfo {year}
  {2014})}\BibitemShut {NoStop}%
\bibitem [{\citenamefont {Reiser}(2007)}]{Reiser2007}%
  \BibitemOpen
  \bibfield  {author} {\bibinfo {author} {\bibfnamefont {M.}~\bibnamefont
  {Reiser}},\ }\href {http://dx.doi.org/10.1002/9783527617623} {\emph {\bibinfo
  {title} {Theory and Design of Charged Particle Beams}}}\ (\bibinfo
  {publisher} {Wiley-VCH Verlag GmbH},\ \bibinfo {year} {2007})\BibitemShut
  {NoStop}%
\bibitem [{\citenamefont {Lawson}(1958)}]{Lawson1958}%
  \BibitemOpen
  \bibfield  {author} {\bibinfo {author} {\bibfnamefont {J.~D.}\ \bibnamefont
  {Lawson}},\ }\bibfield  {booktitle} {\emph {\bibinfo {booktitle} {Journal of
  Electronics and Control}},\ }\href {\doibase 10.1080/00207215808953898}
  {\bibfield  {journal} {\bibinfo  {journal} {J. Electr. Contr.}\ }\textbf
  {\bibinfo {volume} {5}},\ \bibinfo {pages} {146} (\bibinfo {year}
  {1958})}\BibitemShut {NoStop}%
\end{thebibliography}%



\end{document}